\documentclass[preprint,aps,superscriptaddress,showkeys,11pt]{revtex4}
\pdfoutput=1
\usepackage{graphicx}
\usepackage{amsmath}
\usepackage{amsfonts}
\usepackage{amssymb}
\usepackage{xcolor, soul}
\usepackage{epstopdf}
\usepackage{float}
\usepackage{subfigure}
\usepackage{hyperref}
\hypersetup{
     colorlinks   = true,
     citecolor    = red,
     linkcolor    = blue,
     urlcolor     = blue,
}
\def\beq{\begin{equation}}
\def\eeq{\end{equation}}
\def\bea{\begin{eqnarray}}
\def\eea{\end{eqnarray}}
\def\be{\begin{equation}}
\def\ee{\end{equation}}

\def\nno{\nonumber}
\def\bse{\begin{subequations}}
\def\ese{\end{subequations}}

\def\tR{\tilde{R}}
\def\tV{\tilde{V}}
\def\tg{\tilde{g}}

\graphicspath{{./figs/}}

\begin{document}

\title{Minimal inflationary cosmologies and constraints on reheating }

\author{Debaprasad Maity}
 \email{debu@iitg.ac.in}
\author{Pankaj Saha}%
 \email{pankaj.saha@iitg.ac.in}
\affiliation{%
Department of Physics, Indian Institute of Technology Guwahati.\\
 Guwahati, Assam, India 
}%

\date{\today}

\begin{abstract}
With the growing consensus on simple power law inflation models not being favored by the PLANCK observation, dynamics for the non-standard form of the inflaton potential gain significant interest in the recent past. In this paper, we analyze in great detail classes of phenomenologically motivated inflationary models with non-polynomial potential which are the generalization of the potential introduced in \cite{mhiggs}. After the end of inflation, inflaton field will coherently oscillate around its minimum. Depending upon the initial amplitude of the oscillation and coupling parameters standard parametric resonance phenomena will occur. Therefore, we will study how the inflationary model parameters play an important role in understanding the resonant structure of our model under study. Subsequently, the universe will go through the perturbative reheating phase. However, without any specific model consideration, we further study the constraints on our models based on model independent reheating constraint analysis.  
\end{abstract}

\maketitle

\tableofcontents

\section{\label{intro}Introduction}
The inflation \cite{guth, linde1, steinhardt} is a model independent mechanism proposed to solve some of the 
outstanding problems in standard Big-Bang cosmology. It is an early exponential expansion phase of our 
universe, which sets the required initial condition for the standard Big-Bang evolution.  
Over the years large number of models have been introduced to realize this mechanism \cite{encyclopedia}, 
and explain the cosmological observations \cite{PLANCK}.
Out of the large number of models, a particularly interesting class of models that have recently been studied is called $\alpha$-attractor\cite{linde}. It has gained significant attentions because it unifies a large number of existing inflationary models. 
In this paper, we will introduce new classes of inflationary models generalizing the model proposed in \cite{mhiggs}. In order to explain the observation, we phenomenologically consider classes of non-polynomial potentials, which
could be derived from a general scalar-tensor theory in certain limit (shown in appendix-A). At this point let us motivate the reader mentioning the important points of our study. 
It is well known that the general power law canonical potentials of the form $V(\phi) \sim |\phi|^n$ 
are not cosmologically viable because of its prediction of large tensor to scalar ratio. 
In addition, because of super-Planckian value of the field excursion, the effective field theory description may 
be invalid. One of our goals in this paper is to circumvent the above mentioned problems in the framework of 
canonical scalar field model. Therefore, we generalize the power law form of the potential to non-polynomial form so that it can fit well with the observation, and also the inflaton assumes sub-Plankian field excursion. 
After the inflation, the inflaton will go through the oscillatory
phase. Initially because of large oscillation amplitude, the inflaton can decays through parametric resonance depending upon the inflaton coupling with the reheating fields. 
Considering a specific model ($n=2$), we figure out the parameter region where broad parametric resonance happens.
Our analysis shows that as we decrease the inflationary energy scale, the instability bands evolve into wider band, thereby, enhance the strength of the resonance. However, the number of stability/instability region decreases with decreasing the scale we introduced in the model. Detail analysis of this phenomena will be done in our future work. After few initial oscillations, resonant decay will naturally reduce the amplitude of the inflaton oscillation significantly.
Therefore, the perturbative reheating starts to play its role till the radiation domination begins. In this paper we will not discuss about the usual perturbative reheating. However, we should mention that to the best of our knowledge detailed analysis of this 
perturbative phase for arbitrary power law inflaton potential has not been done. We defer this studies for our future publication. However, what we have done instead is the model independent reheating constraint analysis based on the works \cite{kamionkowski,debuGB}, and understand the possible constraints on the model for the successful reheating to be realized.

We structured our paper as follows: In section-\ref{model}, we generalize the model introduced in
\cite{mhiggs}, and study in detail the cosmological dynamics of inflaton starting from inflation to reheating. 
We compute important cosmological parameters such as scalar spectral index $(n_s)$,
the tensor to scalar ratio $(r)$, and the spectral running $(dn_s^k)$ and fit with the experimental observations.
From those cosmological observations, we constrain the parameters of our model. 
After the end of inflation, the inflaton starts to have coherent oscillation around the minimum of the potential, during which the universe will undergo reheating phase.
We also compute the effective equation of state of the oscillating inflaton for our subsequent studies.
In section-\ref{reheating}, we will discuss about how the new scale $\phi_*$ controls the resonance structure during during the first few oscillation of the inflation field. This is very important for pre-heating phenomena. For this we will only consider one single model with $n=2$.
Further detail of this pre-heating phase will be discussed elsewhere.
In section-\ref{reheatingprediction}, we have done the model independent reheating constraint analysis considering the important connection between the end of reheating and the cosmic microwave background (CMB) anisotropy. Finally we concluded and discussed about
our future work.
\section{\label{model}The Model}
\begin{figure}[t]
\centering
\includegraphics[scale=0.8]{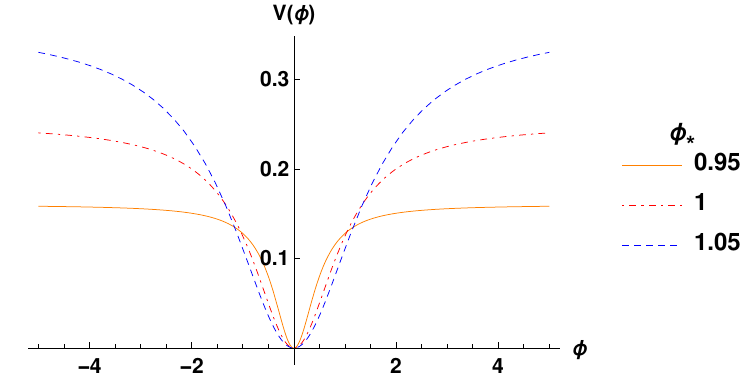}
\caption{\scriptsize An illustration of the dependence of the shape of the potential with the scale $\phi_*$ for $n=2,\lambda=1$ when the two class of the model considered became identical. As we decrease $\phi_*$, the CMB normalization changes the parameter $m$ such that the height as well as the width of the potential decreases. This fact will result in decrease in the field excursion as well as the scalar to tensor ratio and also have significant effect on the post-inflationary dynamics.}
\label{potential}
\end{figure}

As we have discussed in the introduction and also tried to construct in the appendix, our starting point in
this section is the non-polynomial potential which has dominant power law behavior around its minimum. Therefore, we will start by considering the following phenomenological forms of the potential,
\bea
V(\phi) =\begin{cases}  \lambda \frac{m^{4-n} \phi^n}{1+\left(\frac{\phi}{\phi_*}\right)^n} \\ 
\lambda \frac{m^{4-n} \phi^n}{\left(1+\left(\frac{\phi}{\phi_*}\right)^2\right)^\frac{n}{2}},
\end{cases}
\eea
In the above form of the potentials, we have introduced two free parameters $(m~or~\lambda,n)$. Where, the parameter $\lambda$
is defined for $n=4$, which has been studied as minimal Higgs inflation in \cite{mhiggs}. For other value of $n$, we can set 
$\lambda =1$. For large value of $\phi_*$ the potential is plateau like and the associated inflationary scale 
is $\Lambda = \lambda m^{4-n} \phi_*^n$. A simple illustration of the above form of the potentials is shown in the fig.\ref{potential}.
Through out the paper, we will refer type-I for the first form and type-II for the 
second form of the potential. One can further generalize our model by considering the potential to be 
dependent only upon the modulus of the inflaton field. Therefore,
all the odd values of $n$ can be included. For the sake of simplicity we will stick to only even values of $n$.
Another simple generalization of our model can be done by defining $V(\phi)^q$ as a new potential. Where, $q$
would a new parameter.

\subsection{\label{background}Background Equations}
\begin{figure}[t!]
 	\includegraphics[height=5cm,width=6cm]{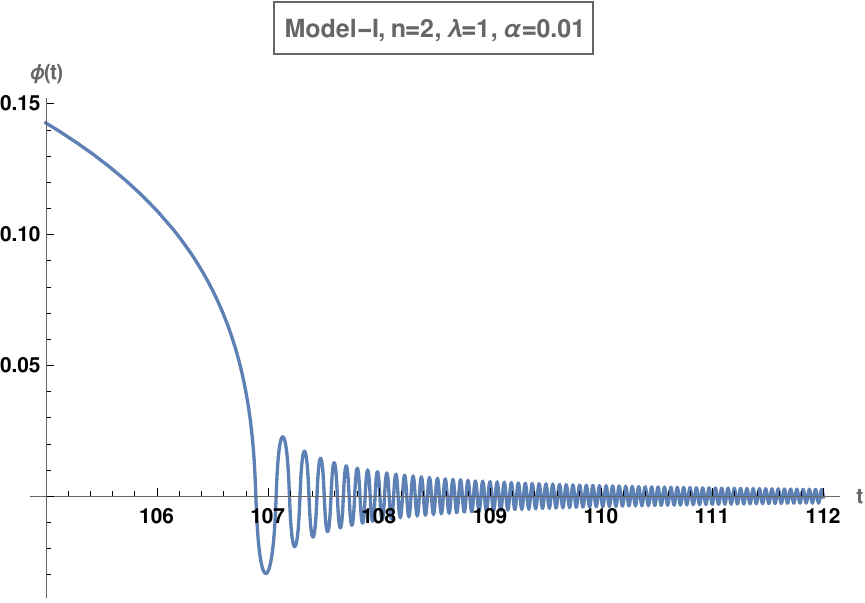}\hspace{0.5cm}\includegraphics[height=5cm,width=6cm]{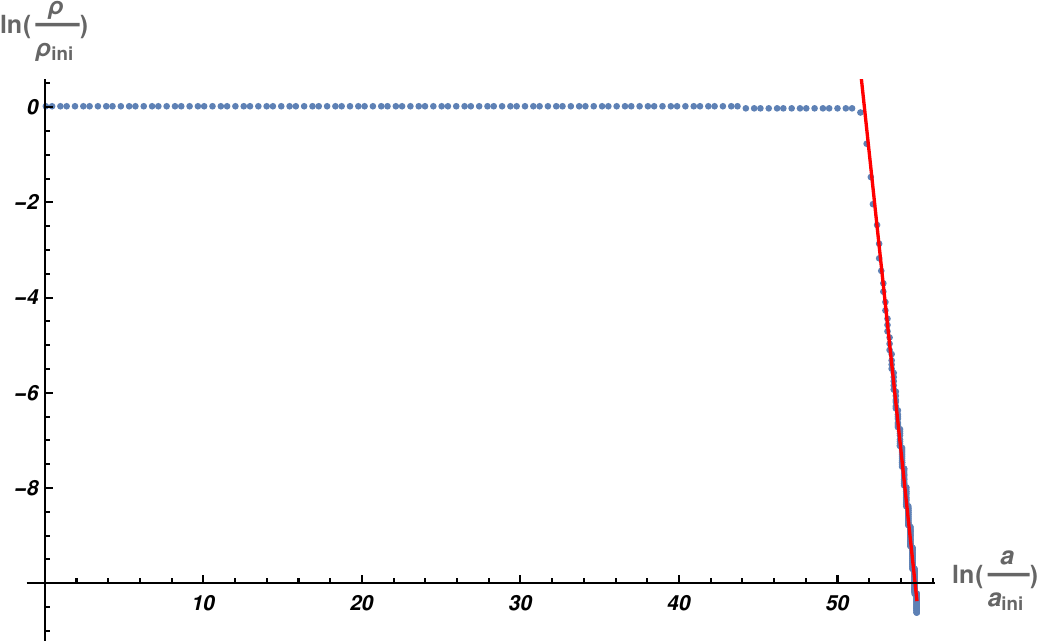}
 	\caption{\scriptsize The evolution of the scalar field with time (in arbitrary unit)and with scale factor for Potential of type I with $n=2$, and $\phi_*=0.01$. time is measure in the unit of $(\frac{m^2}{\phi_*})$}
 	\label{MI-rhoVaplot}
 \end{figure}

In this section we will study in detail the background dynamics using the above form of the potentials.
We will start with the following action,
\bea
S ~=~ \int d^4x \sqrt{-g} \left[\frac {M_p^2}{2}  R - \frac{1}{2}\partial_{\mu}\phi \partial^{\mu}\phi- V(\phi) \right]
\label{action}
\eea
Where $M_p = \frac{1}{\sqrt{8 \pi G}}$ is the reduced Planck mass. 
Assuming the usual Friedmann-Robertson-Walker(FRW) background ansatz for the space-time 
\be
ds^2 = dt^2 - a(t)^2 (dx^2 +dy^2 + dz^2),
\label{frw}
\ee
the system of equations governing the dynamics of inflaton and scale factor are
\bea
3M_{p}^2 H^2= \frac{1}{2}\dot{\phi}^2 + V(\phi)\\
2M_{p}^2 \dot{H} = -\dot{\phi}^2\\
\ddot{\phi} + 3 H \dot{\phi} + V'(\phi).
\label{friedman}
\eea
Where, the usual definition of Hubble constant is $H = {\dot{a}}/a$.
As we have seen, our potential is asymptotically flat for large field value compared to $\phi_*$. This
is condition which is required for the inflationary dynamics is automatically satisfied.
The flatness conditions for the potential during inflation are written in terms of the slow-roll parameters, which are defied as 
 \bea
   \epsilon \equiv \frac{M_p^2}{2} \left( \frac{V'}{V}\right)^2 &=&
   \nno
    \begin{cases}
      \frac{n^2 M_p^2 \phi_* ^{2 n}}{2 \phi ^2 \left(\phi_* ^n+\phi ^n\right)^2} \\
     \frac{\phi_* ^4 n^2 M_p^2}{2 \phi ^2 \left(\phi_* ^2 +\phi ^2\right)^2}
              \end{cases}\\        
               \eta \equiv M_p^2 \left( \frac{V''}{V}\right) &=&
      \begin{cases}
          \frac{n M_p^2 \phi_* ^n \left((n-1) \phi_* ^n-(n+1) \phi ^n\right)}{\phi ^2 \left(\phi_* ^n+\phi ^n\right)^2}\\
        \frac{\phi_* ^2 n M_p^2 \left(\phi_* ^2 (n-1)-3 \phi ^2\right)}{\phi ^2 \left(\phi_* ^2+\phi ^2\right)^2} .
        \end{cases}
  \label{eq-slow-roll}
\eea
During inflation $\epsilon \ll 1$ and $|\eta| \ll 1$. Therefore, the end of inflation 
is usually set by the condition $\epsilon=1$. Let us also define a higher order slow-roll parameter 
related to the third derivative of the potential for spectral running. The expression for the higher order
slow-roll parameter is as follows:
\be
\xi \equiv M_p^4 \left(\frac{V' V'''}{V^2} \right) =
\begin{cases}
\frac{n^2 M_p^4 \phi_* ^{2 n} \left(\left(n^2-3 n+2\right) \phi_* ^{2 n}-4 \left(n^2-1\right) \phi_* ^n 
\phi ^n+\left(n^2+3 n+2\right) \phi ^{2 n}\right)}{\phi ^4 \left(\phi_* ^n+\phi ^n\right)^4}\\
\frac{\phi_* ^4 n^2 M_p^4 \left(\phi_* ^4 \left(n^2-3 n+2\right)+3 \phi_* ^2 (2-3 n) \phi ^2+12 \phi ^4\right)}{\phi ^4 \left(\phi_* ^2+\phi ^2\right)^4} .
\end{cases}
\ee
In addition to provide the successful inflation, all the aforementioned slow-roll 
parameters play very important role in controlling the dynamics of cosmological perturbation during inflation.
An important cosmological parameter which quantifies the amount of inflation is called e-folding number $(N)$, which plays 
crucial role in solving the horizon and flatness problem of standard Big-Bang.  
The e-folding number is expressed as as 
\bea
N = ln\left(\frac{a_{end}}{a_{in}}\right) = \int\limits_{a_{in}}^{a_{end}} d\hspace{0.1cm} lna = \int\limits_{t_{in}}^{t_{end}} H dt   \simeq 
 \int\limits_{\phi_{in}}^{\phi_{end}} \frac{1}{\sqrt{2 \epsilon}} \frac{|d \phi|}{M_p} .
\label{efold}
\eea
 As we have mentioned the inflation ends when $\epsilon=1$, and one can use the eq.(\ref{efold}) to find the 
 value of the inflaton at the beginning of the inflation. 
By solving the aforementioned condition, we can express the e-folding number $N$ into the following form,  
    \begin{equation}
       N=
       \begin{cases}
          \frac{\phi_*^2}{n M_p^2}\left[\frac{1}{(n+2)}(\tilde{\phi}^{(n+2)}-\tilde{\phi}_{end}^{(n+2)}) + \frac{1}{2} (\tilde{\phi}^2-\tilde{\phi}_{end}^2)\right] &  \simeq \frac{\phi_*^2}{n M_p^2} \frac{1}{(n+2)} \tilde{\phi}^{(n+2)} \\
         \frac{\phi_*^2}{n M_p^2} \left[\frac{1}{4} (\tilde{\phi}^4- \tilde{\phi}_{end}^4) + \frac{1}{2} (\tilde{\phi}^2- \tilde{\phi}_{end}^2)\right] & \simeq \frac{\phi_*^2}{4 n M_p^2} \tilde{\phi}^4 .
       \end{cases}
       \label{efold3}
     \end{equation}
Where we have defined, $\tilde{\phi} = {\phi}/{\phi_*}$. In the above expressions for $N$, we have
ignored the contribution coming from $\phi_{end}$, and also the squared term. 
We have numerically checked the validity of those expressions for a wide range
of value of $\phi* \leq {\cal O}(M_p)$. From cosmological observations one needs $N \simeq 50-60$, such that 
the scales of our interest in CMB were in causal contact before the inflation. By using the above mentioned boundary
conditions for the inflaton we have solved for the homogeneous part of inflaton $\phi(t) $ and the scale factor $a(t)$. 
One particular solution has been given in fig.\ref{MI-rhoVaplot}, with
a specific value of the efolding number. Next we study the perturbation around inflationary background 
and derive the relevant cosmological parameters associated the various correlation functions of fluctuation.    

\subsection{\label{pert}  Computation of $(n_s,r,dn_s^k)$}
 As we described in the introduction, the very idea of inflation was introduced to solve 
 some outstanding problems of standard Big-Bang cosmology. Soon it was realized that inflation also provides seed for the large-scale 
 structure of our universe through quantum fluctuation. All the cosmologically relevant inflationary 
observables are identified with various correlation functions 
of those primordial fluctuations calculated in the framework of quantum field theory. 
We have curvature and tensor perturbation. The two and higher point correlation functions of those fluctuation are parametrized by 
power spectrum(see, \cite{cpt1,cpt2,baumann-tasi} for a comprehensive review of Cosmological Perturbation Theory). The scalar curvature power spectrum is given by
 \be
 \mathcal{P_{\mathcal{R}}} = \frac{1}{8 \pi^2} \frac{1}{\epsilon} \frac{H^2}{M_p^2} \bigg|_{k=aH} = \frac{1}{12 \pi^2} \frac{V^3}{M_p^6 (V')^2} .
 \ee
Once, we know the power spectrum, cosmological quantity of our interests are the spectral tilt and its running. 
During inflation a particular inflaton field value corresponds to a particular
momentum mode exiting the horizon. Hence by using the following relation to the leading order
in slow-roll parameters, $\frac{d}{d \hspace{0.1cm} lnk} = \frac{\dot{\phi}}{H}\frac{d}{d\phi}$, one 
obtains the following inflationary observables,
\bea
n_s-1 \equiv \frac{d\hspace{0.1cm} ln \mathcal{P_{\mathcal{R}}}}{d \hspace{0.1cm} lnk} = -6\epsilon + 2 \eta
\eea
\bea
dn_s^k \equiv \frac{dn}{d\hspace{0.1cm} lnk} = -2 \xi + 16 \epsilon \eta - 24 \epsilon^2 .
\eea
Similarly we can compute the tensor power spectrum $\mathcal{P}_{T}$ for the gauge invariant tenor perturbation $h_{ij}$.
To quantify this, standard practice is to define tensor-to-scalar ratio
\bea
r = \frac{\mathcal{P}_{T}}{\mathcal{P}_{R}} = 16\epsilon .
\eea
Once we have all the expression for cosmological quantities in terms of slow roll parameters, by using
eqs.(\ref{eq-slow-roll},\ref{efold3}), and considering $\phi_* \leq {\cal{O}}(1)$ in unit of $M_p$, we express 
$(n_s,r,dn_s^k)$ in terms of $n$,$N$ and $\phi_*$, as 
 \bea \label{nsrvsN}
 1 - n_s &=& 
 \begin{cases}
 	\label{ns-r-ns}
 \frac{2(n+1)}{(n+2)} \frac{1}{N}\\
 \frac{3}{2N}  
 \end{cases} ~~;~~dn_s^k = \begin{cases}
-\frac{(2+3n+n^2)}{(n+2)^2} \frac{1}{N^2}\\
-\frac{3}{4 N^2}
\end{cases}
\\ \nno
  r &=& 
  \begin{cases}
  8n^2 \left(\frac{\phi_*}{M_p}\right)^{\frac{2n}{(n+2)}} \frac{1}{[n(n+2)]^{\frac{2(n+1)}{(n+2)}} N^{\frac{2(n+1)}{(n+2)}}}\\
  \frac{\phi_*}{M_p}  \frac{n^{\frac{1}{2}}}{N^{\frac{3}{2}}}
  \end{cases}
  \label{ns-r-R}
\eea

\begin{figure}[h!]
\centering
\subfigure[ $n_s$ vs $log_{10}\phi_*$ for Model-I]{\includegraphics[width=8.00cm,height=4.00cm]{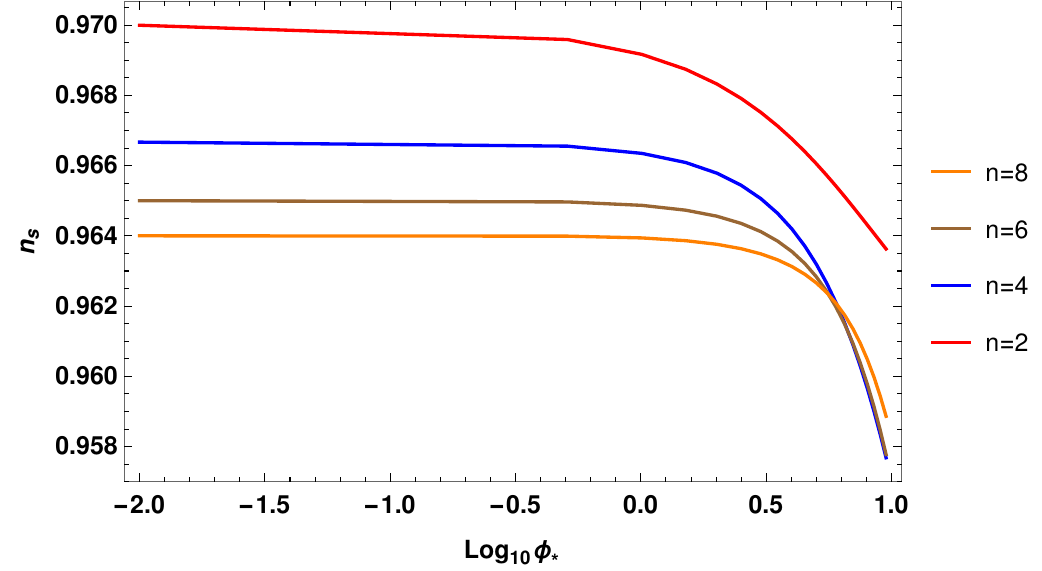}}
\subfigure[ $r$ vs $log_{10}\phi_*$ for Model-I]{\includegraphics[width=8.00cm,height=4.00cm]{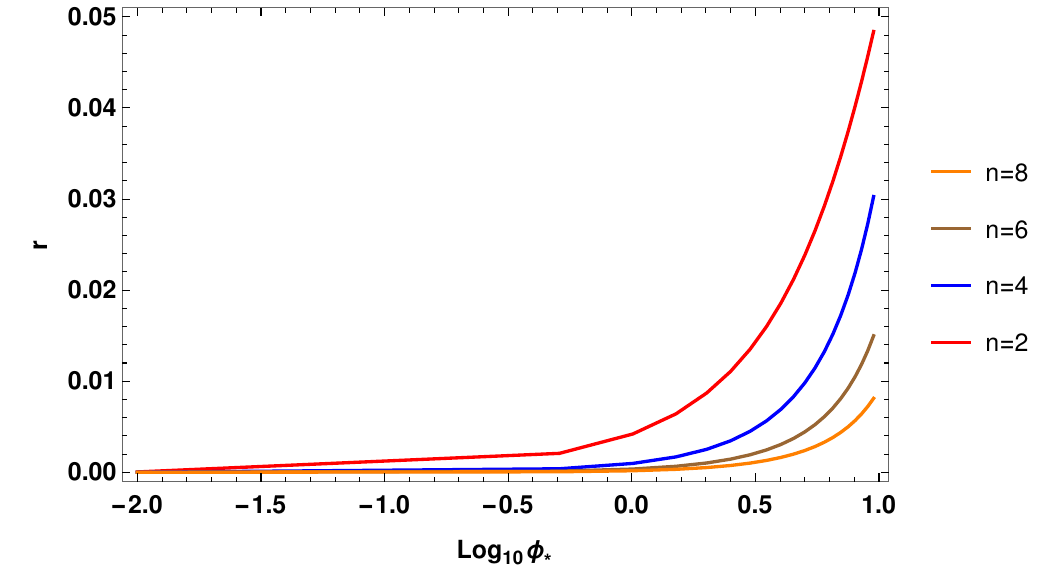}}
\caption{\scriptsize The dependence of $n_s$ and $r$ on the scale $\phi_*$ for fixed number of efolding $(N=50)$ for Potential of type-I.}
\label{fig-nsVlogalpha}
\end{figure}

\begin{figure}[h!]
\centering
\subfigure[ $n_s$ vs $log_{10}\phi_*$ for Model-II]{\includegraphics[width=8.00cm,height=4.00cm]{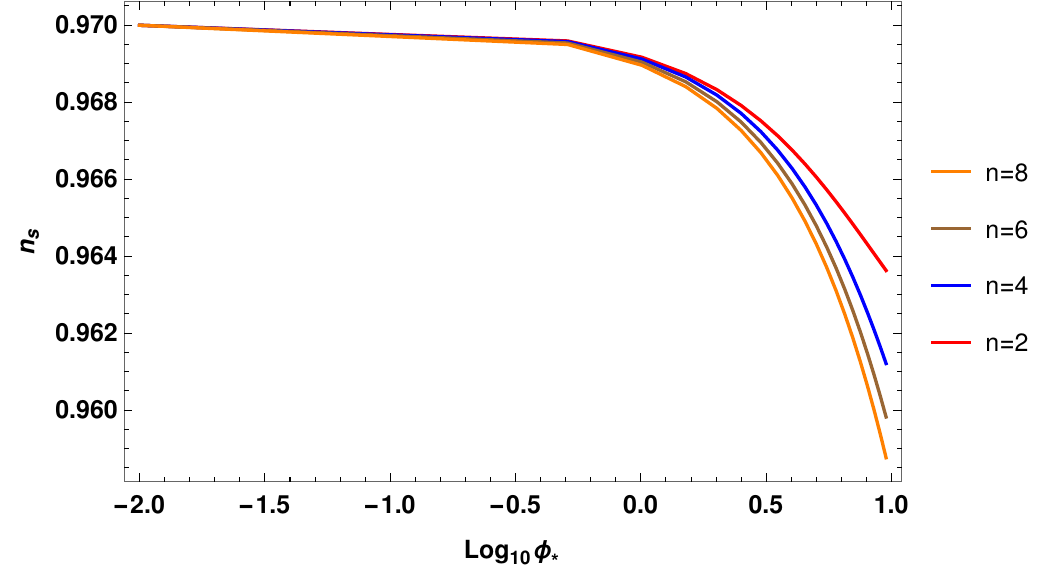}}
\subfigure[ $r$ vs $log_{10}\phi_*$ for Model-II]{\includegraphics[width=8.00cm,height=4.00cm]{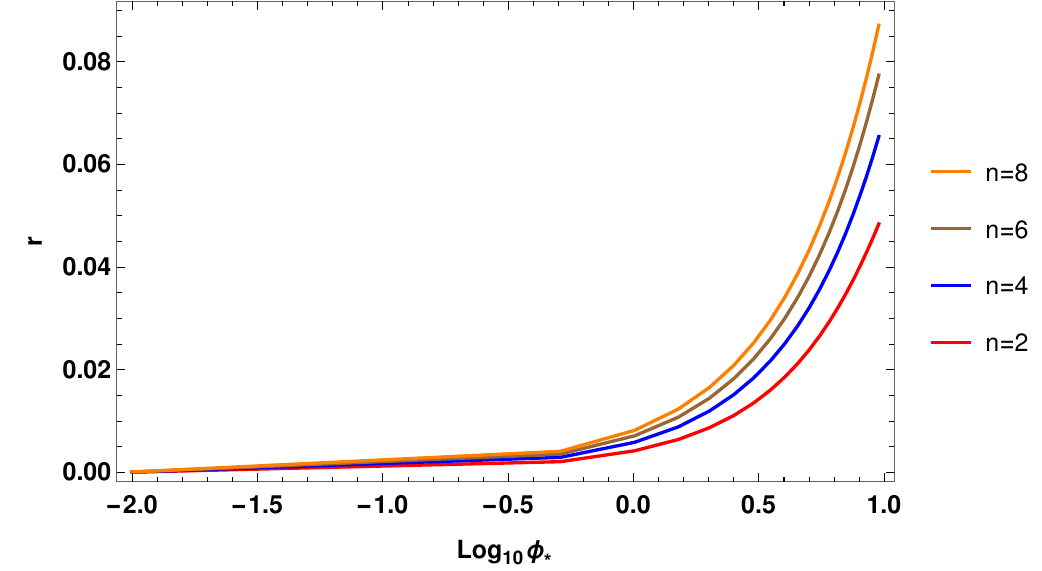}}
\caption{\scriptsize The dependence of $n_s$ and $r$ on the scale $\phi_*$ for fixed number of efolding $(N=50)$ for Potential of type-II}
\label{fig-rVlogalpha}
\end{figure}

At this point, we want to emphasize the fact that the above expansions 
for all the spectral quantities in large-$N$ limit may not always be valid for all inflaton field values as
has been pointed out recently in \cite{Martin:2016iqo}. This fact is indeed true if we look at the 
figs.(\ref{fig-nsVlogalpha},\ref{fig-rVlogalpha}), where, values of $(n_s,r)$ are deviating from the analytic expressions
eq.\ref{nsrvsN} for large $\phi_* > 1~ M_p$. Therefore, for our model, above expansion in large-$N$ for $(n_s,r,dn^k_s)$ are valid only 
in the regime of small-field inflation. Along the line of argument provided in \cite{Martin:2016iqo}, 
we have analytically shown our claim for $n=2$ in  appendix-B.

\begin{figure}[h!]
\centering
\subfigure[Asymptotic behavoir of $n_s$ for model I]{\includegraphics[scale=0.45]{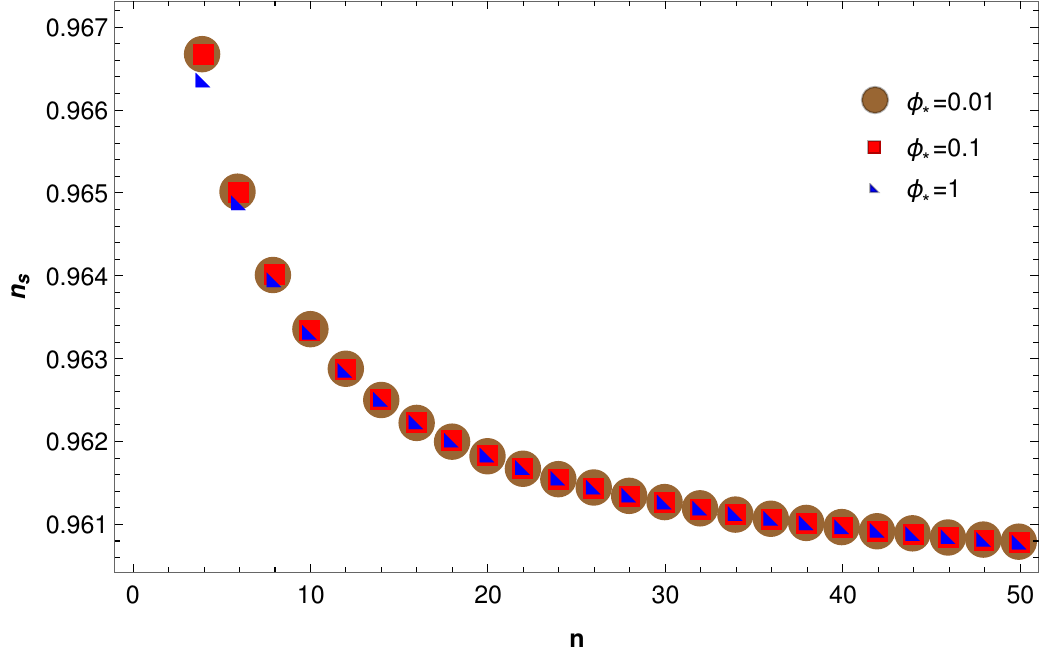}}
\subfigure[Asymptotic Behavior of r for Model II]{\includegraphics[scale=0.45]{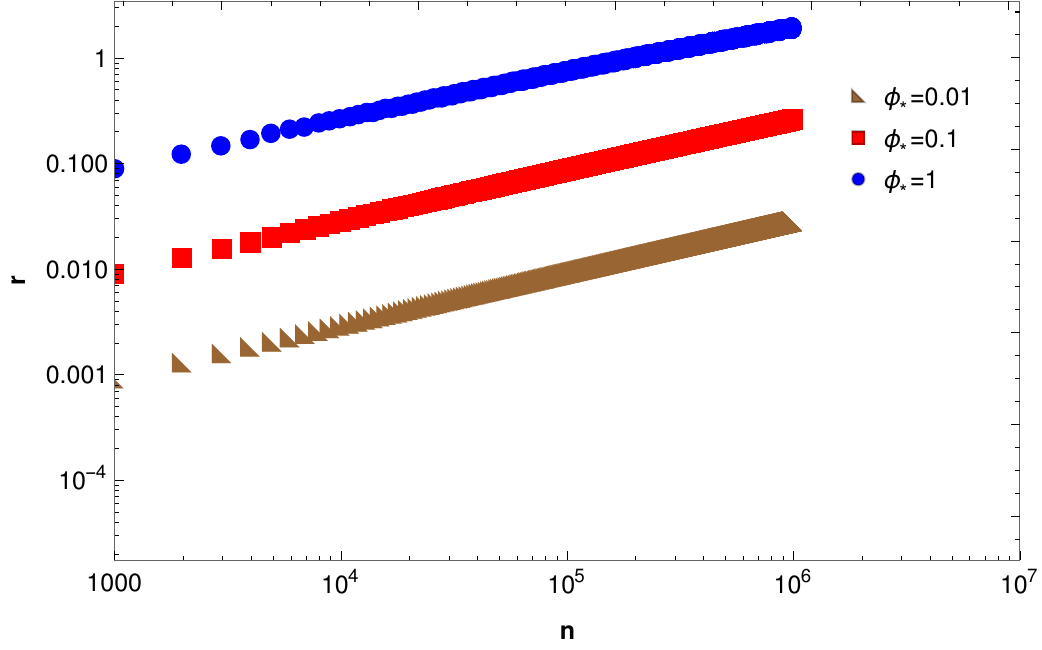}}
\caption{\scriptsize Variation of $n_s$ with $n$ for Model-I with $n=2,4,6,...$, as 
we increase $n$, $n_s \to 0.96$ and variation of $r$ with $n$ for Model-II, it is evident that the value of
$r$ will satisfy the PLANCK bound for $\phi_* \ll 1$}
\label{fig-nsVn}
\end{figure}

From the above analytic expressions for $(n_s,r,dn_s^k)$, some important observations are as follows:
For type-II class of models, we see that the value of $(n_s,dn_s^k)$ are insensitive to the 
value of $n$. This fact can also bee seen from the fig.(\ref{fig-rVlogalpha}),
where, no approximation has been made. In particular one observes that 
for $\phi_* \gtrsim 10 M_p$, value of $(n_s,r)$ start to deviate form each other for different values of $n$.
However, in general tensor to scalar ratio behaves as $r \propto n^{(1/2)}$ to the leading order in $N$. Therefore, form the PLANCK observation
considering the upper bound $r<0.07$, we can put the upper limit on $n$ for a fixed value of $\phi*$. 
From the current observations, it turns out to be very difficult to uniquely fix the form of the potential. Therefore, we need more
theoretical inputs to figure out the full form of the potential. For type-I models, all quantities 
are dependent on $n$, which can also be seen from fig.(\ref{fig-nsVlogalpha}). However, an important fact emerges 
in the limit $n \to \infty$ for Type-II potentials. 
If we take $n \to \infty$ limit, the expressions of $(n_s, r, dn_s^k)$ reduce to
 \bea
 1-n_s \to \frac{2}{N};~~~~ r \to 8 \left(\frac{\phi_*}{M_p}\right)^2 \frac{1}{N^2}~~;~~dn_s^k = - \frac{1}{N^2},
 \eea
which can be identified as a particular model within the class of recently proposed `$\alpha$-attractor'\cite{linde}.
We have numerically checked the aforementioned asymptotic limit of $n_s$ in terms of  $n$, as can be seen in fig.(\ref{fig-nsVn}).
As one observes, from eq.(\ref{ns-r-R}), for $N=50$, the scalar spectral index $n_s \to 0.96$ as $n \to \infty$,
which is the central value of PLANCK observation. 
Therefore, for a wide range of parameter values, we have infinite possible models corresponding to $n =2,3,4,5...$, which 
can successfully explain cosmological observations made by PLANCK \cite{PLANCK}. 
From the field theory point of view, UV completion of our model is an important issue. 
Specifically the supergravity formulation of those form of the potential could be an important direction to study.
 We defer this for our future studies. 

\begin{figure}[h!]
\centering
\subfigure[Model I]{\includegraphics[scale=0.6]{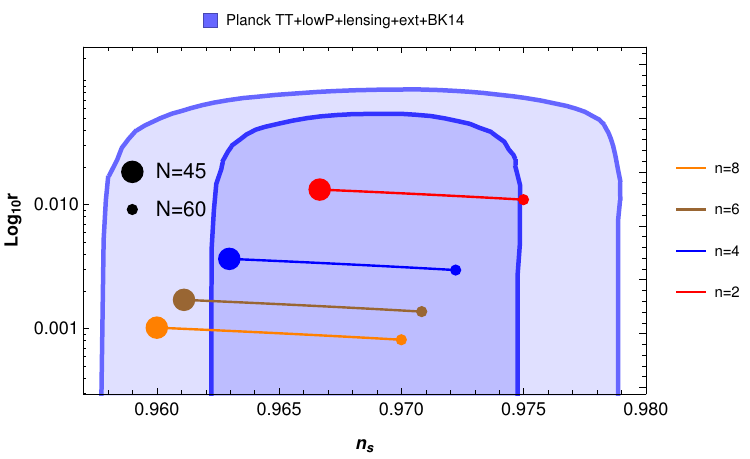}}
\subfigure[Model II]{\includegraphics[scale=0.6]{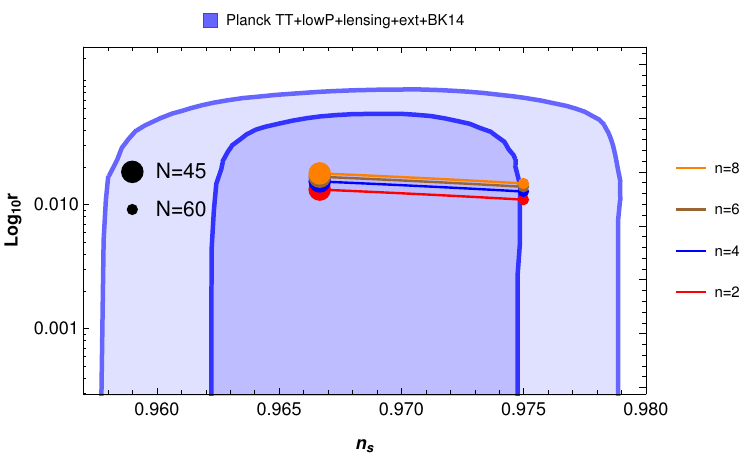}}
\caption{\scriptsize Plot of $n_s$ vs $r$ when $\phi_* =0.1M_p$ for the two potential plotted on Planck 2015 background, as we have seen in our calculation that, for the second type of potential, the calculated quantities are largely independent of $n$. While for the first potential type the change of $n$ has significant effect on the spectral quantities.}
\label{fig-logrVns}
\end{figure}
\begin{table}[h!]
\begin{tabular}{|p{0.7cm}|p{0.6cm}|p{1.0cm}|p{1.5cm}| p{1.5cm} |p{1.0cm}||p{1.0cm}|p{1.5cm}| p{1.5cm} |p{1.0cm}|}
\hline
$\frac{\phi_*}{M_p}$& $n$  &\multicolumn{4}{c|}{Model I}    &\multicolumn{4}{c|}{Model II}  \\
\cline{3-10}
 & & $n_s$ &  $r$ & $dn_s^k$  &  $\Delta \phi $ & $n_s$ & $r$ & $dn_s^k$  &  $\Delta \phi $\\
\hline
\hline
 0.01    &  2  & 0.969 & $4\times10^{-5}$ & -0.00066 & 0.39 & 0.969 & $4\times10^{-5}$ & -0.00066 & 0.39\\
\cline{2-10}
     &  4  & 0.966 &  $2\times10^{-6}$ & -0.00066 & 0.12 & 0.969 & $5\times10^{-5}$ & -0.00060 &  0.47 \\
\cline{2-10}
     &  6  & 0.965 & $3\times10^{-7}$ & -0.00069 & 0.06 & 0.969 & $7\times10^{-5}$ & -0.00060 &  0.51 \\
\cline{2-10}
     &  8  & 0.964 & $1\times10^{-7}$ & -0.00070 & 0.04 & 0.969 & $8 \times 10^{-5}$ &-0.00060 & 0.55  \\
\hline
\hline
 1    &  2  & 0.969 & $4\times10^{-3}$ & -0.0006 & 3.53 & 0.969 & $4\times10^{-3}$ & -0.0006 & 3.53 \\
\cline{2-10}
     &  4  & 0.966 & $9.6\times10^{-4}$ & -0.0007 & 2.13 & 0.969 & $6\times10^{-3}$ & -0.0006 &  4.0 \\  
\cline{2-10}
     &  6  & 0.964 & $3.5\times10^{-4}$ & -0.0007 & 1.47 & 0.969 & $7\times10^{-3}$ & -0.0006 &  4.3 \\
\cline{2-10}
     &  8  & 0.964 & $1.7\times10^{-4}$ & -0.0007 & 1.1 & 0.969 & $8\times10^{-3}$ & -0.0006 & 4.7  \\
\hline
\hline
\end{tabular}
\caption{\scriptsize The spectral quantities for different 
values of $n$ for 50 efolding. The two values of $\phi_*$ are chosen to illustrate that we can have both small field and large field inflation 
depending on the value of $\phi_*$. The general trend for the variation of these quantities with $\phi_*$ is illustrated in the figures (\ref{fig-nsVlogalpha}-\ref{fig-rVlogalpha}) }
\label{nsrtable}
\end{table}

So far all we have discussed is directly related to the cosmological observation made by PLANCK. 
Another important quantity of theoretical interest we would like to compute is Lyth bound \cite{lythbound} $\Delta \phi$. 
This quantity measures the difference of field values which is traversed by the inflaton
field during inflation. This is so calculated that for a particular model $\Delta \phi$ is the maximum possible
value for a particular efolding number. Inflation is a semi-classical phenomena. It is believed that natural cut off scale for any theory
minimally or non-minimally coupled with gravity is Planck scale $M_p$. Therefore, amount of inflaton field value 
can naturally be a good measure to tell us the effective validity of a model under study in the effective field theory
language. Hence the calculated expression for the field excursion in terms of $N$ and $\phi_*$ are:
\bea
\Delta \phi \gtrsim M_p N \sqrt{\frac{r}{8}} = \begin{cases}
M_p \left(  \frac{n}{n+2} \right) \frac{1}{\left[n(n+2)\right]^{\left(\frac{n+1}{n+2}\right)}} N^{\frac{1}{(n+2)}}\\
\frac{M_p}{2} \left(\frac{\phi_*}{M_P}\right)^{\frac{1}{2}} N^{\frac{1}{4}}
\end{cases}
\eea
All the quantities we have discussed so far is independent of $m$ or $\lambda$. (At this point let us again 
remind the reader that for $n \neq 4$, $\lambda$ is a dimensionless quartic coupling parameter. While for $n\neq 4$, $m$ is 
dimensionful parameter, and we set $\lambda =1$).
However, comparing the inflationary power spectrum with the PLANCK normalization we will determine the value of
$m$ or $\lambda$ and then calculate all the other quantities of our interest. 
The expression for the power spectrum of the curvature perturbation is 
\bea
\mathcal{P_R} =
\begin{cases}
\frac{\lambda}{12 \pi^2 n^2} \left(\frac{m}{M_p}\right)^{4-n} \left(\frac{\phi_*}{M_p}\right)^{\frac{n^2}{n+2}} \left[n(n+2) N\right]^{\frac{2(n+1)}{(n+2)}}\\
\frac{2}{3 \pi^2} \frac{\lambda}{\sqrt{n}} \left(\frac{m}{M_p}\right)^{(4-n)} \left(\frac{\phi_*}{M_p}\right)^{(n-1)} N^{\frac{3}{2}}
\end{cases} = 2.4 \times 10^{-9}.
\label{powspectrum}
\eea 
As mentioned we considered the PLANCK normalization: $\mathcal{P}_R$ at the pivot scale $k/a_0 = 0.05 Mpc^{-1}$,
and corresponding estimated scalar spectral index is $n_s = 0.9682 \pm 0.0062$. 

After having all our necessary expressions for all the cosmological quantities, we have plotted our main results in $(n_s,r)$
space and compared with the experimental values $n_s=0.968 \pm 0.006$ and upper limit on $r<0.11$ in fig.(\ref{fig-logrVns}). In the 
table-(\ref{nsrtable}), we have given some sample values of all the cosmologically relevant quantities for different values
of theoretical parameters. As we have mentioned already, we found infinitely many model potentials with a universal
shape. Most interesting case would probably
be for $n=4$. In the recent paper \cite{mhiggs}, it has been identified as a minimal Higgs inflation. 
Of course this identification is not straight forward. However for small field value, we can 
certainly Taylor expand the potential, and identify the coupling $\lambda$ as Higgs quartic coupling which can be set to its electroweak value. 
However, renormalization analysis needs to be done in order to do this identification. 

At this point let us re-emphasize the fact that the observation made by PLANCK strongly disfavors the usual power law inflation 
with $n\geq 2$. In this paper we showed that problems of those power law inflationary models can be cured with a non-polynomial generalization of the potential. 
We plotted the dependence of the $(n_s, r)$ on the inflationary energy scale $\phi*$ in figs.(\ref{fig-nsVlogalpha}-\ref{fig-rVlogalpha}).
For both type of models, it is clearly matching with our approximate analytic expression eq.(\ref{ns-r-R}). In the subsequent section we will see how the reheating prediction will constrain
the value of $N$ depending upon the reheating
temperature consistent with PLANCK.

 \subsection{End of inflation and general equation of state}
 In this section we will be interested in the dynamics of the inflaton field
 after the inflation. During this phase the inflaton field oscillates coherently around the minimum
 of the potential. At the beginning the oscillation dynamics will naturally be dependent upon 
 the inflation scale $\phi_*$ because of the large amplitude. This is the stage during which non-perturbative
 particle production will be effective. Therefore, resonant particle production will take place and
 conversion of energy from the inflaton to matter particles will be highly efficient. This phenomena 
 is usually known as pre-heating of the universe. In this section we will discuss about
 the late time behaviour of the inflaton, specifically focusing on the dynamics of the
 energy density of the inflaton field. After the many oscillations, 
 when the amplitude of the inflaton decreases much below the $\phi_*$, the dynamics will be
 controlled by usual power law potential. As we have emphasized the coherent oscillation is 
 very important in standard treatment of reheating. 
 For any models of inflation this is thought to be an important criteria to have successful reheating.  
 In this section, we will first discuss the evolution of inflaton and its energy density 
 in full generality for all classes of potentials. As mentioned before, at late time the potential can be approximated as
 \be
 V(\phi) = \lambda m^{4-n} \phi^n .
 \ee
 
 \begin{table}[t!]
 	\begin{tabular}{|p{0.6cm}|p{2.0cm}|p{2.5cm}|p{1.5cm}| p{2.0cm} |p{2.0cm}|}
 		\hline
 		$n$& $ \textit{w}   =\frac{n-2}{n+2} $ & $p = 3(1 + \textit{w})$&\multicolumn{2}{c|}{ p from fitting}\\
 		\cline{4-5}
 		&&&Model I&  Model II\\
 		\hline
 		\hline
 		$2$ & 0 & 3 & 3.12 & 3.12\\
 		\hline
 		4 & $\frac{1}{3}$ & 4 & 3.99 & 3.93\\
 		\hline
 		6 & $\frac{1}{2}$ & 4.5 & 4.56 & 4.45\\
 		\hline
 		8 & $\frac{3}{5}$ & 4.8 & 4.83 & 4.74 \\
 		\hline
 	\end{tabular}
 	\caption{The variation of inflation energy density with scale factor for various potential}
 	\label{T-rhoVa}
 \end{table}
 
 In cosmology for any dynamical field such as inflaton, one usually defines the equation of state parameter $\textit{w}$.
 For the oscillating inflaton, when the time scale of oscillation about the minimum of a potential is small enough compared
 to the background expansion time scale, by using virial theorem effective equation of state for a potential of the 
 form $V(\phi) \propto \phi^n$ can be expressed as\cite{mukhanov.book}
 \be
 \textit{w} \equiv \frac{P_{\phi}}{\rho_{\phi}} \simeq \frac{\langle \phi V'(\phi) \rangle - \langle 2 V\rangle}{\langle \phi V'(\phi)\rangle + \langle 2 V\rangle}=\frac{n-2}{n+2} .
 \label{w-n}
 \ee
 Therefore, in an expanding background, the evolution of energy density $\rho_{\phi}$ of the inflaton averaged over many oscillation
 will follow,
 \bea
 \dot{\rho}_{\phi} + 3 H (1+\textit{w}) \rho_{\phi} = 0 .
 \eea
 At late time we relate the energy density$(\rho_{\phi})$ of the universe
 (assuming that the universe is dominated by a single component) and the scale factor $(a)$ as
 \be
 \rho_{\phi} \propto a^{-3(1+\textit{w})} = a^{-p} .
 \label{rho-n}
 \ee
 In the table-\ref{T-rhoVa}, we provide some theoretical as well as numerically fitting values corresponding to 
 the equation of state parameter $\textit{w}$ of the inflaton and the power law evolution of the energy 
 density namely the value of $p$.   
 
 In the following sections will be considering those equation of state parameters 
 and study their role in the subsequent cosmological evolution. We will first
 discuss about the constraint on reheating phenomena by taking the model
 independent approach, where explicit dynamics during reheating phase
 will not be considered.          

   \section{\label{reheating}Pre-heating: parametric resonance and their dependence on inflation scale $\phi_*$}
   Reheating is an important phase of the early universe, when all the matter field is assumed
   to be produced from the decay of inflaton. Initial study on this mechanism
   was based on the perturbative quantum field theory
   \cite{Albrecht:1982mp,Dolgov:1982th,Abbott:1982hn,Traschen:1990sw}. 
   However, it was soon realized that this approach may not be efficient enough for successful reheating.
   In general reheating phenomena is a complicated non-linear
   dynamics of inflaton coupled with matter fields at finite temperature and the process of their thermalization. In the seminal work by Kofman, Linde and Starobinsky\cite{Kofman:1994rk,Kofman:1997yn}
   (see also\cite{Shtanov:1994ce}), the idea of non-perturbative resonant production of 
   particles has been introduced\cite{Landau:mec}. Though the full non-linear theory 
   of preheating is still not well understood but a significant advancement in this field has
   been made and lot of works are going on    \cite{Bassett:2005xm,Allahverdi:2010xz,Amin:2014eta}. It is generally believed that the 
   reheating phase usually happens in two stages. In the first stage, particle production is due to  parametric resonance known as `preheating' followed by the perturbative reheating.  
   
   Therefore, in this section we discuss about non-perturbative particle production via parametric resonance phenomena for a specific model ($n=2$). It is evident that initial few oscillations after the end
   of inflation play important role at this stage. Therefore, we will see how the inflationary scale $\phi_*$ which controls the shape of
   the potential, effects the structure of the resonance for reheating fields. Hence, through resonance structure, we may be able to further restrict the parameter space of our model. We defer the detailed of this resonance 
   phenomena for our future study. 
   
     The inflaton field oscillating coherently after inflation acts as a \textit{classical external force} leading to the production and growth of quantum boson fields via the bose condensation. We write the Lagrangian for the daughter scalar field $\chi$ as.
   \be
   \mathcal{L}_{\chi} = \frac 1 2 \partial_{\mu} \chi \partial^{\mu} \chi - \frac 1 2 m_{\chi}^2 \chi^2 - \frac{1}{2}g^2 \phi^2 \chi^2.
   \ee
   Where, $m_{\chi}$ is the mass of the $\chi$ particle. The matter field $\chi$ satisfies the following equation:
   \bea
   {\ddot {\chi}} + 3 H \dot{\chi} -\frac {1}{a^2} \nabla^2 \chi  +(m_{\chi}^2 + g^2 \phi^2)  \chi = 0 .
   \eea 
   Decomposing the scalar field operator into Fourier modes,
   \bea
   \chi(t,x) = \int \frac {d^3 {\bf k}}{(2 \pi)^{2/3}} [a_{\bf k} ~\chi_k(t) e^{ i {\bf k}\cdot{\bf x}} + {a}_{\bf k}\dagger ~\chi_k(t) e^{- i {\bf k}\cdot{\bf x}}],
   \eea
   the mode equation for $\chi_k(t)$ takes the following from 
   \bea \label{chieq}
   \ddot{\chi}_k + 3 H \dot{\chi}_k + \left(\frac {k^2}{a^2} + m_{\chi}^2 + g^2 \phi^2 \right) \chi_k = 0 .
   \eea 
   Where, $a_{\bf k},{a}_{\bf k}\dagger$ are the creation and annihilation operators respectively.
   The parametric resonance phenomena with periodic background force can be best explained
   though the stability/instability diagram arising from the above generalized Mathieu  
   equation\cite{McLachlan, MW}. To study this in the expanding cosmological background we rescale the field 
   variable $\chi_k$ and the inflaton field $\phi$ as follows 
   \bea
   \label{rescale}
   \nno
   \chi_k \to a^{-\frac{3}{2}} X_k ~~;~~ \phi \to a^{- \frac{3}{2}} \Phi .
   \eea
   Where we have set the initial value of the scale factor to unity at the point from where we started our numerical computation. 
   With the above rescaling, the mode Eq.(\ref{chieq}) now turns out to be
   \be
   X_k'' + \omega^2_k X_k = 0
   \label{chieq2}
   \ee
   where
   $$\omega^2_k \equiv \frac{k^2}{a^2 m^2} + \frac{g^2 \phi^2_0}{m^2 a^3} \Phi^2(t) + \Delta~, ~~~~~~~~~~ \Delta \equiv -\frac{3}{4}(3H^2 + 2\dot{H}),$$
   and ``prime" is taken with respect to rescaled dimensionless time variable $z=mt$. 
   As we are only considering $n=2$, soon after the inflation ends, the background evolution approximately satisfies
   $H^2 \equiv \dot{H} \ll m^2$. Therefore, we set $\Delta \simeq 0$ in the above eq.\ref{chieq} to simplify our computation. 
   Due to rescaling, the rescaled background inflaton field oscillates with almost constant amplitude. 
   We set $\phi_0$ as the initial amplitude of the coherent inflaton oscillation. 
   
   If we ignore the expansion of the universe, eq.(\ref{chieq2}) can be identified as a Hill's differential equation 
   with parameters $\kappa= \frac{k^2}{m^2 a^2}$ and $q = \frac{g^2 \phi_0^2}{m^2 a^3}$. The solution of this equation is known to exhibit parametric resonance depending on the value of the parameters $(q, \kappa)$. 
   If the value of the parameters $(q, \kappa)$ is within certain `instability bands' the solution of the eq.(\ref{chieq2}) grows exponentially as 
   $X_k \propto exp(\mu_k z)$. Where, Floquet exponent, $\mu_k$, parametrizes the strength of the resonance. Therefore, depending upon
   the value of $\mu_k$, corresponding $\chi$-particle of momentum $k$ will grow exponentially. This indefinite growth of any mode is just the consequence of neglecting the expansion of the universe as well as the back-reaction of the produced $\chi$ particles. When the expansion is included the parameters in the Hill's equation becomes time dependent. However, it can be seen that the relative change in $q$ during  oscillation is
   \be
   \frac{1}{m} \frac{\dot{q}}{q} = -\frac{3H}{m}.
   \ee
   During reheating period, $H \ll m$, hence, $q$ parameter can be taken as constant.
   
\begin{figure}[t]
   	\centering
   	\includegraphics[scale=0.4]{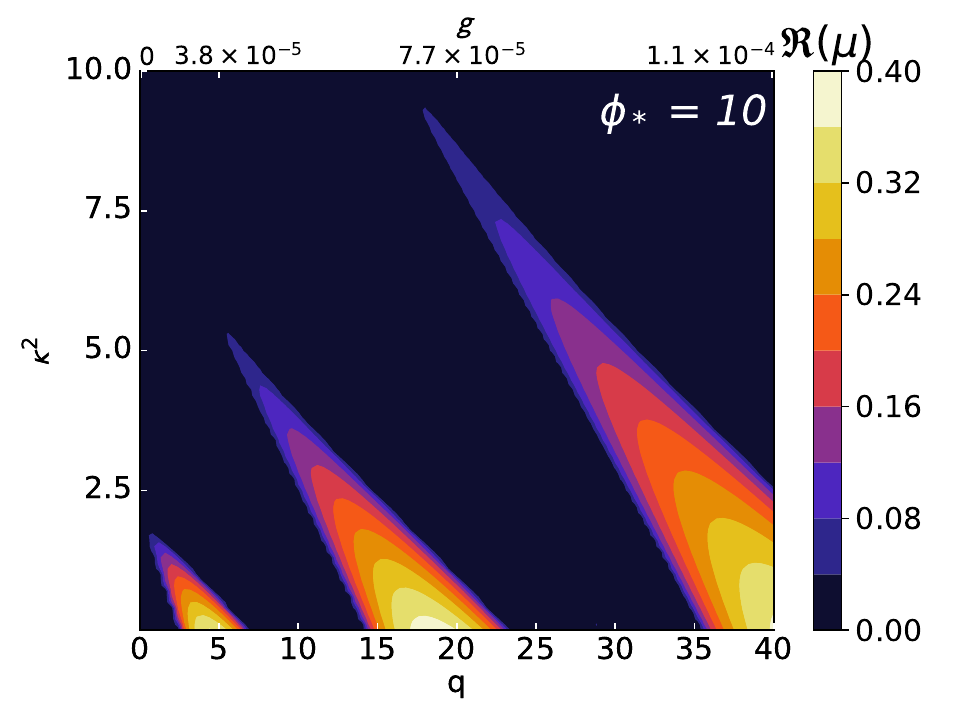}~~ \includegraphics[scale=0.4]{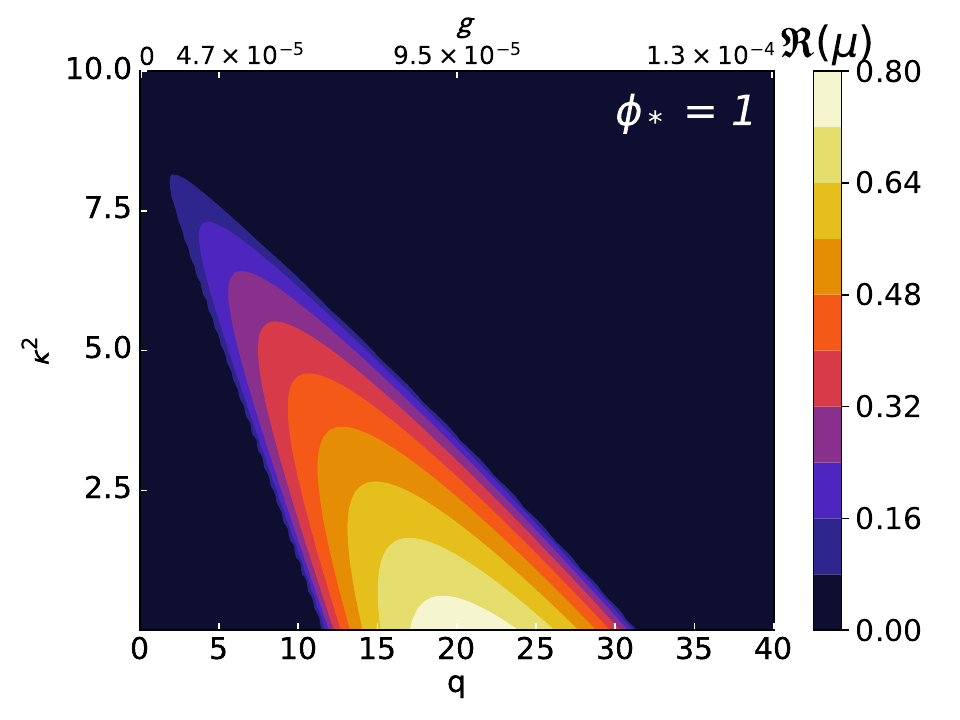} \\
   	\includegraphics[scale=0.4]{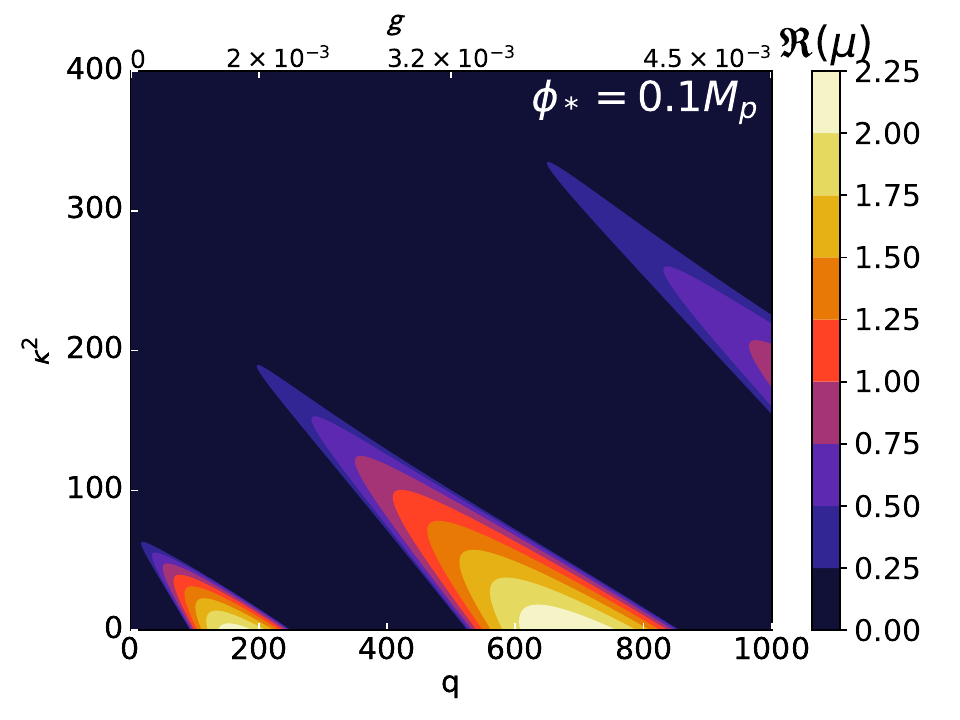}~~\includegraphics[scale=0.4]{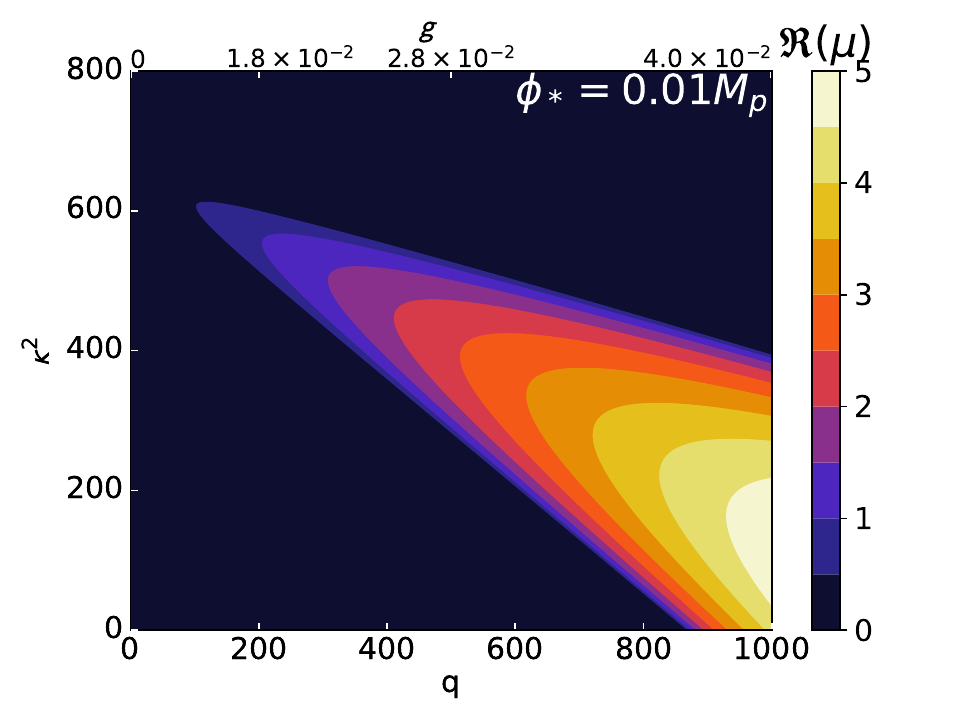}
   	\caption{\scriptsize Stability/instability charts for the non-perturbative production of reheating field $\chi$ (Eq.(\ref{chieq}) in $(\frac{k^2}{m^2}, q)$ space.}
   	\label{stability}
   \end{figure}

\begin{figure}[h!]
   	\centering
   	\subfigure[]{\includegraphics[scale=0.5]{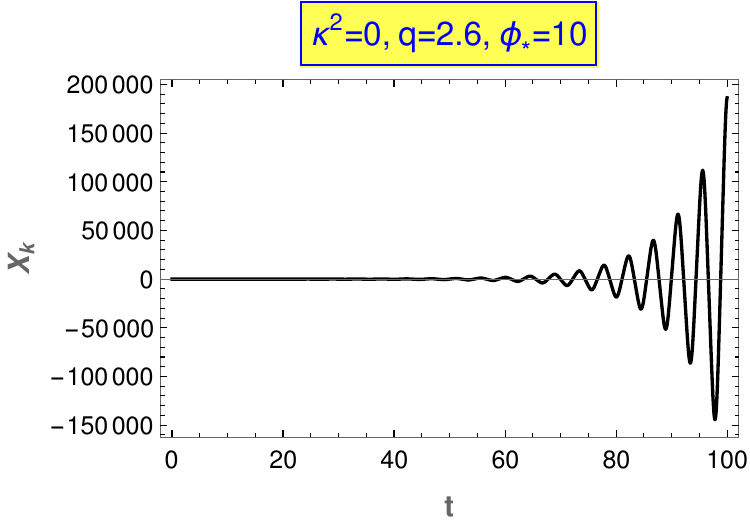}}
   	\subfigure[]{\includegraphics[scale=0.5]{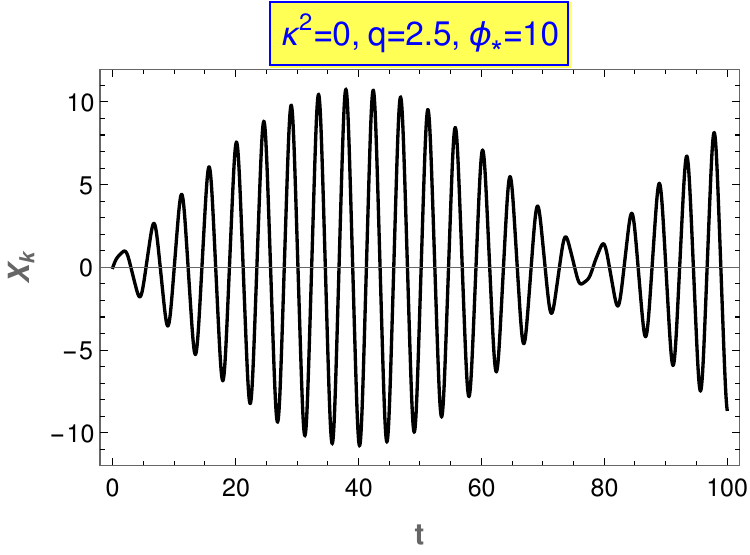}}
   	\caption{\scriptsize The zero mode of the produced field ( $ X_k = a^{\frac{3}{2}} \chi$) for values of slightly inside (a) and outside (b) of the instability band.}
   	\label{chi-n2-10}
   \end{figure}

   The structure of resonance for chaotic inflationary model has been well studied.
   As has been mentioned, the scale $\phi_*$ plays very important role during inflation. Therefore, 
   our main goal would be to understand the role of $\phi_*$ on the structure of resonance. 
   To compare the effect for different $\phi_*$ we consider $n=2$, and measure time in unit of $m_{10}$ which is the value of $m$ corresponding to $\phi_* = 10M_p$. We present the contour plots for the Floquet exponent 
   in $(q, \kappa)$ space for different values of $\phi_*$ as shown in fig.\ref{stability}. 
   We also show how the zero mode function $(X_{k=0})$ grows for $\phi_* =10$ in fig.\ref{chi-n2-10} for two values of $q$, one taken from inside and another just outside the unstable region.	
   The stability/instability chart has been computed for the first oscillation taking $a(t_0) =1$. Therefore,
   with time the magnitude of $(q,\kappa)$ parameters decrease which measures duration
   of preheating period. 
   
   Now we will examine the effects of $\phi_*$ on the structure of resonance. The very first point that we would like
   to point out is the dependence of the Floquet exponent on inflation scale $\phi_*$. We can clearly see that as we decrease
   $\phi_*$ from $(10, 1, 0.1, 0.01)$ in unit of Planck, the maximum value of Floquet exponent increases as $(0.4, 0.8, 2.25, 5)$ respectively.
   Therefore, with decreasing value of $\phi_*$, the resonance becomes stronger and simultaneously the band width also 
   increases. This can also be seen from the approximate analytic expression for the band width
   \bea
   \Delta k \simeq \left(\frac{m}{m_{10}}\right)^{\frac 1 2} (0.4 g \dot{\phi})^{\frac 1 2}.
   \eea 
   Where, velocity of $\phi$ field is measured in unit of $m^{-1}$. However, in order to compare the results for different $\phi_*$,
   we set a fixed time scale $m_{10}$ corresponding to $\phi_*=10$.
   Hence, as we decrease $\phi_*$, $m$ decreases from CMB normalization, which enhance the frequency of 
   the inflaton oscillation and consequently the resonance band width becomes wider. 
   Moreover, for large $\phi_*$, long wavelength modes of $\chi$ field needs stronger coupling to get excited. 
   From the effective field theory point of view, small scale inflation suggests the parameter should be $\phi_* < 1$ in unit of Planck mass. Further, the broad parametric resonance constrains the coupling parameter $g \gtrsim (10^{-3},10^{-2})$ for $ \phi_* \simeq (0.1, 0.01) M_p$.
   This is in sharp contrast with the usual chaotic inflation, where inflationary observables do not have much effect on the 
   reheating coupling parameter $g$, and consequently the reheating temperature if we consider the shifted minimum of the inflaton 
   potential. We will do detail lattice study on this issue in the subsequent publication.
   
   
   To this end, let us mentioned an important point which we will defer for our future studies. As we decrease the value of $\phi_*$ below $1 M_p$, the effective mass $m_{eff}^2 = V''(\phi)$ of the inflation field becomes negative in certain range of inflaton field values after the end of inflation and remains so for first few oscillations. This will lead to tachyonic preheating and will have important consequences specifically 
   with regard to the gravitational production. We will differ detailed study on this issue for our future work.

\section{\label{reheatingprediction} Model independent constraints from reheating predictions}
  After inflation, reheating is the most important phase, where, all the visible matter energy
  will be pumped in. In this section, we will try to constrain our model parameters 
  without any specific mechanism of reheating. The background
evolution of cosmological scales from inflation to the present day and the conservation of entropy density
provide us important constraints on reheating as well as our model parameters.  
Reheating is the supposed to be the integral part of the inflationary paradigm. 
However, because of the single observable universe, it 
is very difficult to understand this process by the present day cosmological observation.
Thermalization process erases all the information about the initial conditions which is the most
important part of this phase. To understand this phase an indirect attempt has been made in the recent past \cite{liddle,kamionkowski,cook} 
through the evolution of cosmological scales and the entropy density by parametrizing 
it by reheating temperature $(T_{re})$, equation of state $(w_{re})$, and efolding 
number $(N_{re})$. In this section we follow the reference \cite{debuGB} by taking into
account the two stage reheating phase generalizing the formalism of \cite{liddle}. 
Our main goal is to understand the possible constraint on our minimal inflationary models. 
As we have seen from previous analysis, all the cosmological quantities during inflation can 
be expressed in terms of two main parameters $(m~or~\lambda, \phi_*)$ for a particular model.  
Because of two stage reheating process, the suitable reheating parameters 
are as follows, $(N_{re}=N^1_{re}+N^2_{re},T_{re},w^1_{re},w^2_{re})$. 
Where, $N^1_{re}, N^2_{re}$ are efolding number during the first and second stage of the reheating 
phase with the equation of state $w^1_{re}, w^2_{re}$ respectively. 
At the initial stage the oscillating inflaton will be the dominant component, and
at the end radiation must be the dominant component. Therefore, instead of 
taking the equation of state as free parameters, we will be considering only the following 
particular case 
\bea
w^1_{re} = \frac{n-2}{n+2} ~~;~~~ w^2_{re} = \frac 1 3 .
\label{eqparameter}
\eea
 We also assumed the change of reheating phase from the first to the 
second stage as instantaneous.


A particular scale $k$ going out of the horizon during inflation will re-enter the horizon during
usual cosmological evolution. This fact will provide us an important 
relation among different phases of expansion parametrizing by enfolding number as follows 
\bea
\ln{\left(\frac k {a_0 H_0}\right)} = \ln{\left(\frac {a_k H_k}{a_0 H_0}\right)} =-N_k -\sum_{i=1}^{2}N^i_{re} -
\ln{\left(\frac {a_{re} H_k}{a_0 H_0}\right)},
\label{scalek}
\eea  
In the above expressions, use has been made of $k = a_0 H_0 = a_k H_k$. 
Where, $(a_{re}, a_0)$ are the cosmological scale factor at the end of the reheating phase and at the present time respectively.
$(N_k,H_k)$ are the efolding number and the Hubble parameter respectively for a particular scale $k$ 
which exits the horizon during inflation. Therefore, following mathematical expressions will be used in the final numerical
calculation, 
\begin{eqnarray}
H_k &=&  \sqrt{\frac{V(\phi_k)}{3 M_p^2}} = \begin{cases} \left(\frac{\lambda \phi_*^n}{3 M_p^2}\right)^{\frac 1 2} 
\frac{m^{\frac{4-n}{2}} \tilde{\phi}_k^{\frac {n}{2}}}
{\left(1+\tilde{\phi}_k^n\right)^{\frac 1 2}} \\ 
 \left(\frac{\lambda \phi_*^n}{3 M_p^2}\right)^{\frac 1 2} \frac{m^{\frac{4-n}{2}} 
 \tilde{\phi}_k^{\frac{n}{2}}}{\left(1+\tilde{\phi}_k^2\right)^{\frac{n}{4}}},
 \end{cases}
\\
N_k &=&  \frac 1 {M_p} 
\int_{\phi_{end}}^{\phi_{k}} \frac {1} {\sqrt{2 \epsilon}} d\phi \simeq
\begin{cases}
          \frac{\phi_*^2}{n M_p^2}\left[\frac{1}{(n+2)}\tilde{\phi}_k^{(n+2)} + \frac{1}{2} \tilde{\phi}_k^2\right] \\
         \frac{\phi_*^2}{n M_p^2} \left[\frac{1}{4} \tilde{\phi}_k^4 + \frac{1}{2} \tilde{\phi}_k^2\right] 
       \end{cases}
       \label{HandN}
\end{eqnarray}

\begin{figure}[t!]
\begin{center}
  \includegraphics[width=006.0cm,height=04.0cm]{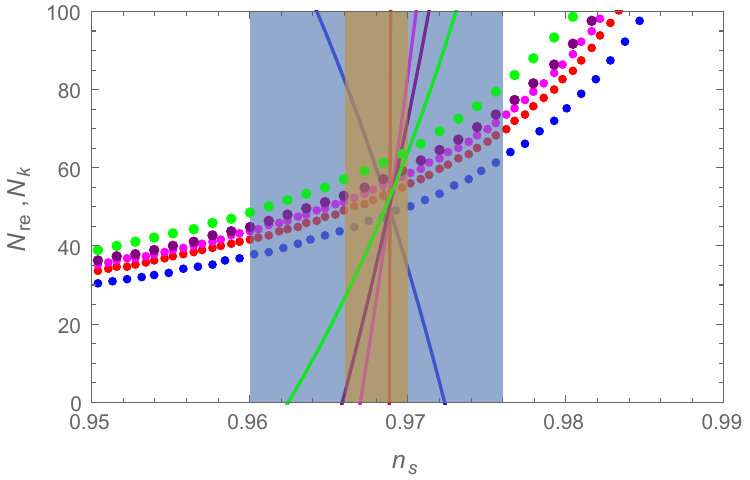}
  \includegraphics[width=006.0cm,height=04.0cm]{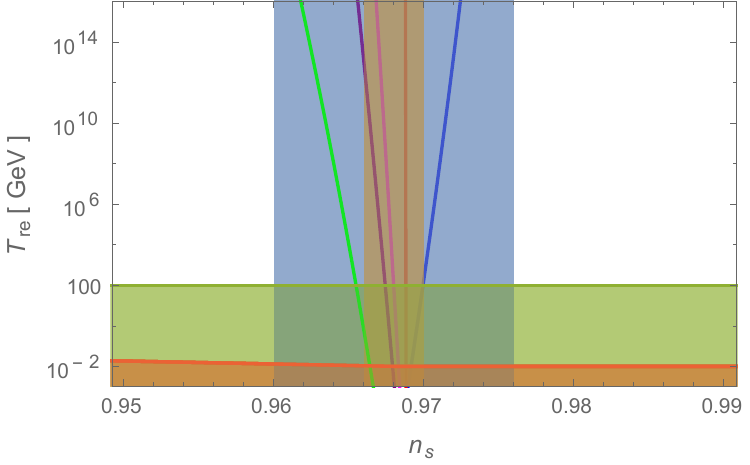}
 \caption{\scriptsize Variation of $(N_{re}(\mbox{solid}), N_{k}(\mbox{dotted}),T_{re})$ as a function of $n_s$ have been plotted for 
$\phi_* = 0.01 M_p$. This is the plot for Model-I. (Blue, red, magenta, brown, green) curves correspond to $n=(2,4,6,8,30)$. 
Each curve corresponds to a specific set of equation of state parameters $(w^1_{re},w^2_{re}) =((n-2)/(n+2),1/3)$ during reheating.
We also consider $N^1_{re}=N^2_{re}$. The light blue shaded region 
corresponds to the $1 \sigma$ bounds on $n_s$ from Planck. The brown shaded region corresponds to the $1 \sigma$
bounds of a further CMB experiment with sensitivity $\pm 10^{-3}$ \cite{limit1,limit2}, using the same central $n_s$ value as
Planck. Temperatures below the horizontal red line is ruled out by BBN. The deep green shaded region is below the 
electroweak scale, assumed 100 GeV for reference.} 
\label{tre1}
\end{center}
\end{figure}

$\phi_k$ and $\phi_{end}$ are the inflaton field values corresponding to a particular scale $k$ crossing the inflationary horizon,
and at the end of inflation respectively.
In the above expressions, we have ignored the contribution coming from the inflaton field value $\phi_{end}$.
It is important to note that, in principle we can write the field value at a particular scale $k$
in terms of $n_s, r$, by inverting those relations. Because of non-linear form, we will numerically solve those.
The above unknown efolding numbers during reheating will certainly be dependent upon the energy densities 
$(\rho_{end},\rho_{re})$, at the end of inflaton (beginning of reheating phase) and at the end of the 
reheating phase( beginning of the standard radiation dominated phase);  
\bea
\ln\left(\frac {\rho_{end}}{\rho_{re}}\right) = 3(1+ w^1_{re}) N^1_{re} + 3(1+ w^2_{re}) N^2_{re}= 3\sum_{i=1}^{2} (1+ w^i_{re}) N^i_{re}.
\label{rho2}
\eea
Above two eqs.(\ref{scalek},\ref{rho2}), can be easily generalized for multi-stage inflation with different 
equation of state parameters. 
As has been mentioned, after the end of reheating standard evolution of our universe is precisely known in terms of
energy density and the equilibrium temperature of the relativistic degrees of freedom such as photon and
the neutrinos. Therefore, the equilibrium temperature after the end of reheating phase, $T_{re}$, is related
to temperature $(T_0, T_{\nu 0})$ of the CMB photon and neutrino background at the present day respectively, as follows
\bea \label{entropy}
g_{re} T_{re}^3 = \left(\frac {a_0}{a_{re}}\right)^3\left( 2 T_0^3 + 6 \frac 7 8 T_{\nu 0}^3\right).
\eea
The basic underlying assumption of the above equation is the conservation of reheating entropy during the
the evolution from the radiation dominated phase to the current phase. $g_{re}$ is the number of
relativistic degrees of freedom after the end of reheating phase. 
We also use the following relation between the two temperatures, $T_{\nu 0} = (4/11)^{1/3} T_0$. 
For further calculation, we define a quantity, $\gamma = N^2_{re}/N^1_{re}$. If we identify
the scale of cosmological importance $k$ as the pivot scale of PLANCK, so that $k/a_0 = 0.05 Mpc^{-1}$, and
the corresponding estimated scalar spectral index $n_s = 0.9682 \pm 0.0062$, we arrive at the following 
equation for the efolding number during reheating period, and the reheating temperature,

\begin{figure}
\begin{center}
  \includegraphics[width=006.0cm,height=04.0cm]{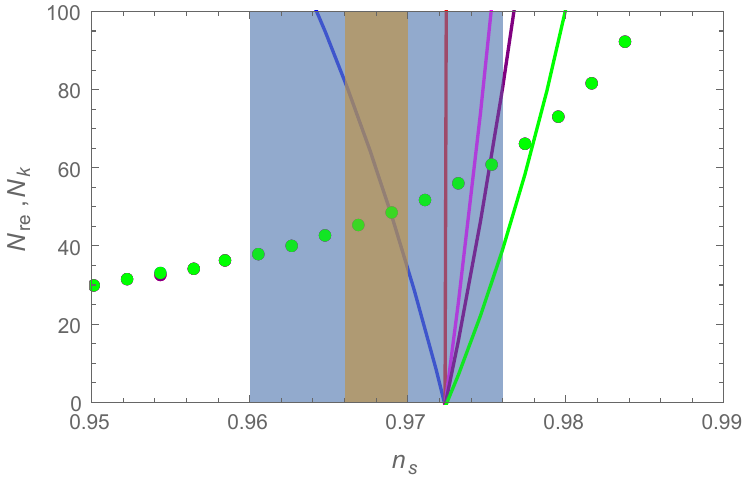}
  \includegraphics[width=006.0cm,height=04.0cm]{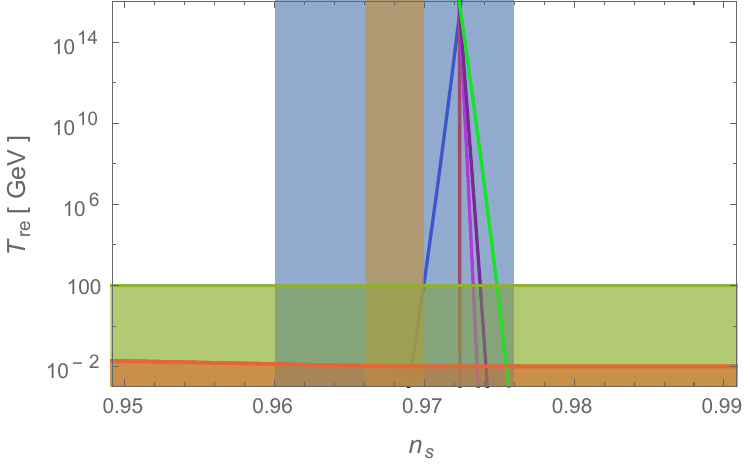}
 \caption{\scriptsize Variation of $(N_{re}(\mbox{solid}), N_{k}(\mbox{dotted}),T_{re})$ as a function of $n_s$ have been plotted for 
$\phi_* = 0.01 M_p$. This is the plot for Model-II. (Blue, red, magenta, brown, green) curves correspond to $n=(2,4,6,8,30)$. 
Each curve corresponds to a specific set of equation of state parameters $(w^1_{re},w^2_{re}) =((n-2)/(n+2),1/3)$ during reheating.
We also consider $N^1_{re}=N^2_{re}$. From the left figure, one clearly sees that the behavior of $N_{k}$ is independent of $n$} 
\label{tre2}
\end{center}
\end{figure}

\begin{eqnarray} 
&& N_{re} = \frac{4(1+\gamma)}{(1-3w_{re1})+\gamma(1-3 w_{re2})}\left[61.6 - \ln\left(\frac{V_{end}^\frac{1}{4}}{H_k}\right) -N_k\right]\\
&& T_{re} = \left[\left(\frac{43}{11g_{re}}\right)^\frac{1}{3} \frac{a_0 T_0}{k} H_ke^{-N_k}\right]^{\frac{3[(1+w_{re1})+\gamma(1+w_{re2})]}
{(3w_{re1}-1)+\gamma(3 w_{re2}-1)}}
 \left[\frac{3^2.5V_{end}}{\pi^2g_{re}}\right]^{\frac{1+\gamma}{(1-3w_{re1})+\gamma(1-3 w_{re2})}}  .
 \label{nretre}
\end{eqnarray}

In the above derivation, we have used $g_{re} = 100$. 
Before discussing any further, let us provide the general descriptions of the figures we have drawn in this section.
As has been mentioned before, we have considered specific values of equation of
state parameter $(w^1_{re},w^2_{re}) =((n-2)/(n+2), 1/3)$ in compatible with our model
discussed in the previous section. Important to 
mention regarding a special point in the aforementioned state space is $(1/3,1/3)$ which is realized for $n=4$. 
Analytically one can check that at this special point both $(T_{re}, N_{re})$ become indeterministic seen in eq.(\ref{nretre}). 
This fact corresponds to all the vertical solid red lines in $(n_s~vs~T_{re})$ and $(n_s~vs~N_{re})$ plots. 
We have considered $\gamma =1$ as our arbitrary choice. Each curve corresponds
to different values of $n$. On the same plot 
of $(n_s~vs~N_{re})$, we also plotted $(n_s~vs~N_k)$ corresponding to the dotted curves for different models.
One particularly notices that for the second type model in fig.(\ref{tre2}), behavior of $(n_s~vs~N_k)$ is same for all different value of
$n$. This universality is inherited from the fact that $n_s$ does not really depend upon $n$. Therefore, background
dynamics because of the second type model for different values of $n$ are almost universal. 
However, prediction of $(T_{re}, N_{re})$ are dependent upon the value of $n$ through
the equation of state parameter eq.(\ref{eqparameter}). 
At this stage, we would like to remind the reader again that for a wide range of $\phi_*$, all the models predict very small
value of tensor to scalar ratio $r$. Therefore, we will be discussing all the constraints without explicitly mentioning $r$. 
Given the overall description of all the plots, we now set to discuss the
prediction and constraints for two different models.  
In the table-(\ref{nsrtable}) we provide the important numbers for reheating temperature and the 
efolding number. We provided only the limiting values of $T_{re}$ which are still allowed from the 
cosmological observation. 
 
\begin{figure}
	\begin{center}
		\includegraphics[width=0012.5cm,height=04.0cm]{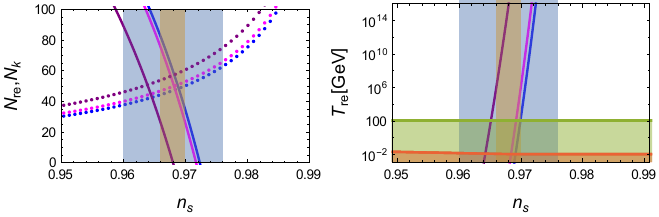}
		\caption{\scriptsize Variation of $(N_{re}(\mbox{solid}), N_{k}(\mbox{dotted}),T_{re})$ as a function of $n_s$ have been plotted for for three different values of $\phi_*$. (Blue, magenta, purple) curves are for $\phi_* = (0.01, 0.1,10) M_p$ respectively. We consider only $n=2$ for Model-I. All the other parameters remain the same as for the previous plots.}		
		\label{tre2phi*n2}
	\end{center}
\end{figure}
 
\begin{table}[h!]
\begin{tabular}{|p{0.6cm}|p{1.2cm}|p{1.5cm}|p{0.7cm} |p{0.7cm}|p{1.2cm}|p{1.5cm}|p{0.7cm} |p{0.7cm}|}
\hline
 $n$  &\multicolumn{4}{c|}{Model-I} &\multicolumn{4}{c|}{Model-II} \\
\cline{2-9}
  & $n_s$ & $T_{re}$(GeV) & $N_{re}$ & $N_{k}$ & $n_s$ & $T_{re}$(GeV) &  $N_{re}$& $N_k$\\
\hline
\hline
 2  & \begin{tabular}{c}0.9723 \\0.9702 \end{tabular} &  \begin{tabular}{c} $1\times10^{15}$\\ $1\times10^{3} $ \end{tabular}&
 \begin{tabular}{c} 0.4\\32\end{tabular} & \begin{tabular}{c}54\\50\end{tabular} 
&\begin{tabular}{c} 0.9723\\0.9702\end{tabular} & \begin{tabular}{c} $2\times10^{13}$\\ $1\times10^{3}$ \end{tabular}&
\begin{tabular}{c}0.4\\32 \end{tabular}& \begin{tabular}{c}  54\\53 \end{tabular}\\
\hline
       6  &\begin{tabular}{c} 0.9670\\0.9679 \end{tabular}&\begin{tabular}{c} $1\times10^{14}$\\ $2\times10^{3} $ 
\end{tabular}& \begin{tabular}{c}00\\23\end{tabular} &\begin{tabular}{c} 53\\55 
\end{tabular}&\begin{tabular}{c} 0.9724\\0.9728 \end{tabular}&\begin{tabular}{c} $4\times10^{14}$\\$4\times10^{3}$ \end{tabular}&
\begin{tabular}{c} 00\\24\end{tabular} &\begin{tabular}{c} 54\\56 \end{tabular} \\
\hline
       8  & \begin{tabular}{c}0.9659\\0.9673 \end{tabular}&\begin{tabular}{c} $7\times10^{13}$\\ $1\times10^{3}$\end{tabular} &\begin{tabular}{c} 
0.6\\23 \end{tabular}&\begin{tabular}{c} 53\\55 \end{tabular}&
\begin{tabular}{c} 0.9725\\0.9730\end{tabular} & \begin{tabular}{c} $3\times10^{14}$\\ $2\times10^{3}$ \end{tabular}&
\begin{tabular}{c} 0.4\\24 \end{tabular}&\begin{tabular}{c} 55\\57 \end{tabular}\\
\hline
       30  & \begin{tabular}{c}0.9625\\0.9653 \end{tabular}& \begin{tabular}{c} $6\times10^{13}$\\ $1\times10^{3} $ \end{tabular}&\begin{tabular}{c} 
00\\21 \end{tabular}&\begin{tabular}{c} 52\\56 \end{tabular}&
\begin{tabular}{c}0.9726\\0.9736\end{tabular} &\begin{tabular}{c} $6\times10^{14}$\\ $1\times10^{3}$ \end{tabular}&
\begin{tabular}{c} 0.5\\23 \end{tabular}&\begin{tabular}{c} 55\\59 \end{tabular} \\
\hline
\end{tabular}
\caption{\scriptsize Some sample values of $(n_s, T_{re}, N_{re},N_k)$ are give for two different models
for $n=(2,6,8,30)$. As we have mentioned, for $n=4$, $(T_{re}, N_{re})$ become indeterministic.
All these prediction are for $\phi_* = 0.01 M_p$.}
\label{nsrtable}
\end{table}

\begin{figure}[t!]
	\begin{center}
			\includegraphics[width=13.0cm,height=04.0cm]{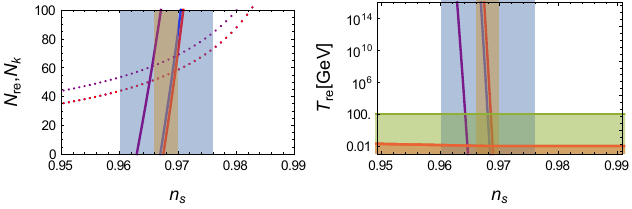}
		\caption{\scriptsize Variation of $(N_{re}(\mbox{solid}), N_{k}(\mbox{dotted}),T_{re})$ as a function of $n_s$ have been plotted for for three different values of $\phi_*$. (Blue, magenta, purple) curves are for $\phi_* = (0.01, 0.1,10) M_p$ respectively. We consider only $n=6$ for Model-I. All the other parameters remain the same as for the previous plots.} 
		\label{tre2phi*n6}
	\end{center}
\end{figure}

From the figure we see that for a very small change in $n_s$, the variation of
reheating temperature is very high. Therefore, reheating temperature provides tight constraints on the possible
values of efolding number $N_{re}$ during reheating phase. Except for $n=4$, if we restrict the value of $T_{re} \gtrsim 10^3$ GeV, 
the efolding number turned out to be $N_{re} \lesssim 35$ during reheating. As an example, for $n=2$, we find 
spectral index lies within $0.9723 \lesssim n_s \lesssim 0.9702$. Within this range of spectral index, the reheating temperature has to be within $1\times10^{15} \gtrsim T_{re} \gtrsim  1\times10^{3}$
in unit of GeV. This restriction in turn fixed 
the possible range of efolding number within a very narrow range $50< N < 54$ for $n=2$. For other value of $n$, the ranges are provided in the table-\ref{nsrtable}. What we can infer from our analysis in this section is that reheating  constraint does not allow $n$ to be vary large specifically for type-I model.
 Whereas for type-II mode, the prediction of $n_s$ is almost independent of $n$ for $\phi_* < M_p$ However in the figs.\ref{tre2phi*n2},\ref{tre2phi*n6},we have plotted the dependence of various reheating parameters for different values of $\phi_*$. We have plotted only for type-I model and $n=2,6$. For all the other models qualitative behaviors of those plots will be same, except $n=4$.

 With increasing value of the equation of state, efolding number during reheating  
$N_{re}$, decreases for a fixed value of reheating temperature. This essentially means that
as one increases the value of inflationary equation of state $w$, faster will be the thermalization process, 
therefore, earlier will be the radiation dominated phase. 
In our subsequent full numerical solutions, we have observed this fact considering the 
evolution of all the important components during reheating.  
We also noted that as we decrease the value of $\gamma$, the prediction of $(T_{re}, N_{re})$ will
be controlled by inflation equation of state parameter $ w^1_{re}$. 

\section{\label{conclusion}Summary and Conclusion}
Before we conclude, let us summarize the main results of our study. As emphasized, we tried to constrain specific classes of inflationary models based on the inflation and dark matter abundance. Effective field theory consideration constraints $\phi_*$ to be less than unity in Planck unit. As a result we have sub-Planckian field excursion during inflation.
In this regime of $\phi_*$ the value of inflationary observables $(n_s,r,dn_s^k)$ saturate to a constant value depending on the e-folding number $N$ and the power law index $n$. Furthermore, requirement of broad parametric resonance
constraints the coupling parameter $g \gtrsim 10^{-3}$ for $n=2, \phi_* = 0.1$. For fixed $n$, if we further reduce the value of $\phi_*$, 
the lower limit on $g$ increases, however, resonance becomes stronger and broader. Therefore, instant transfer of energy from inflaton to reheating field is possible within few oscillations of the inflaton field. Detailed analysis of this issue will be done in our subsequent publication. 

 In the first part of this paper, we have constructed two new classes of inflationary model
with non-polynomial modification of the inflaton potential. In the appendix we have tried to
construct such potential from a most general non-minimal scalar tensor theory. 
In certain region of the parameter space, our models coincide with the aforementioned 
scalar-tensor theory. It would be interesting to construct such potential from more fundamental 
approach. Interesting property of
these classes of potentials is that they have infinitely large flat plateau.
Therefore, the inflation can be naturally realized because of this shift symmetry.
As a result, the predictions of the models for inflationary observables are not very much 
sensitive to the detail form specifically near the minimum of the potential. 
Importantly our model fits extremely well with latest cosmological observation made by PLANCK. 
All the necessary scales assume below Planck scale value, which may imply that our model predictions are robust
against quantum correction. 
Detailed computation on the ultra-violate effect on our model could be important and
we left it for our future work. Depending upon the choice of scale, 
in our model we realize both large field as well as small field inflation.
However, for both the cases, the prediction
of tensor to scalar ratio $(r)$ turned out to be significantly small. In the end we have studied model independent reheating constraint analysis and discuss about further constraint coming from the reheating when connecting with the CMB anisotropy. 

Another important aspect of our model is that we can have significantly low inflationary 
Hubble scale $H_*$ unlike the usual power law inflationary model. The value could be 
as low as $\sim 10^9$ GeV for different values of $n = 4 ,6, 8$ for $\phi_*\simeq 0.0001 M_p$.
It is well known that low value of $H_*$ could be interesting in the context of 
Higgs vacuum instability. As has been pointed out in \cite{1407.3141, 1506.04065}, 
during as well as after the inflation the quantum fluctuation of Higgs field can destabilize the standard model metastable Higgs vacuum 
at around $\Lambda_I =10^{11}$ GeV. 
However, this instability crucially depends upon the value of $H_*$, 
and also the height of the Higgs potential. Therefore, comparing the naive scale
dependence between $\Lambda_I$ and $H^*$, our model have 
potential to save the Higgs vacuum from decaying into the global vacuum. We leave 
the detailed study on this issue for our future work.

  
\section{Acknowledgement}
We would like to thank our HEP and Gravity group members for their valuable comments and discussions.

\appendix

\section{Towards derivation of our model potential}

  In this section starting from non-minimal scalar-tensor theory,
we will try to construct our model potentials which were a priori ad hoc in nature. 
As is well known, inflationary models based on power law potential $V(\phi) \sim \phi^n$ are simple but have been ruled out in general 
because of their large prediction of tensor to scalar ratio. Moreover, the models with large plateaus (Starobinsky or $\alpha$-attractors) 
are found to be most favored form the PLANCK observation. While most of these plateau models can be cast into exponential potential, 
plateau potentials with power-law form have also been discussed in super gravity\cite{Dimopoulos:2014boa, Dimopoulos:2016zhy} and non-minimal coupling to gravity\cite{Eshaghi:2015rta, Broy:2016rfg}. In this section we will try to construct our model based on this non-minimally coupled scalar-tensor theory. 
We will see, how simple power-law potentials in the Jordan frame can give rise to the plateau potentials of 
desired form in the Einstein frame. However, this transformed models coincide with our minimal models 
only in a limiting regime (weak conformal coupling). At this point
let us point out that equivalence between the Einstein frame and Jordon frame is an important 
question to ask. This issue has been discussed \cite{Futamase:1987ua, Kaiser:1994vs, Hwang:1996np, Deruelle:2010ht, Postma:2014vaa, Banerjee:2016, Bhattacharya:2017pqc}, from theoretical as well as cosmological point of views. 

Nevertheless, our motivation in this section is to construct our desired form of the potentials which 
we have shown to be in different class of models rather tan $\alpha$ attractor model. 
We start with the following non-minimally coupled scalar-tensor 
theory, 
\be
S_J = \int d^4x \sqrt{-g} \left[ \frac{\Omega(\varphi)}{2} M_p^2 R - \frac{\omega(\varphi)}{2} g^{\mu \nu } \partial_{\mu}\varphi \partial_{\nu}\varphi - V(\varphi)     \right],
\ee
where, $\Omega(\varphi), \omega(\varphi)$ are arbitrary function of a scalar field $\varphi$. We will chose a specific form of those
function for our later purpose.
To get the action in the Einstein frame, one performs the following conformal transformation as,
\be
\tg_{\mu \nu} = \Omega(\varphi) g_{\mu \nu} ,
\ee
The action in the Einstein frame can be written as\cite{Fujii:2003pa}
\be
S_E = \int d^4x \sqrt{-\tg} \left[\frac{M_p^2}{2} \tR -\frac{1}{2} F^2(\varphi) \tg^{\mu \nu} \partial_{\mu}\varphi \partial_{\nu}\varphi - \tV(\varphi)      \right]
\ee
	Where, we have assumed that $\omega(\varphi) = \Omega(\varphi)$ and $F$ and the new potential can be found to be,
\be
F^2(\varphi) = \frac{3 M_p^2}{2} \frac{\Omega'^2(\varphi)}{\Omega^2(\varphi)} + 1 ~~;~~\tV(\varphi) = \frac{V(\varphi)}{\Omega^2(\varphi)}
\label{Funct}
\ee
Now, we choose the following non-minimal coupling function \cite{Kallosh:2013tua, Galante:2014ifa}, for 
$\Omega^2(\varphi)$, 
\bea
   \Omega^2(\varphi) =
   \begin{cases}
   	1+ \xi (\frac{\varphi}{M_p})^n\\
   	\left[1+ \xi (\frac{\varphi}{M_p})^2 \right]^{\frac{n}{2}} .
   \end{cases}
   \label{omega}
   \eea
Therefore, applying (\ref{omega}), we find F and $\tV$ as,
   \bea
   F^2(\varphi) = 
   \begin{cases}
   \frac{3 n^2 \xi ^2 \left(\frac{\varphi }{M_p}\right)^{2 (n-1)}}{8 \left[1 + \xi  \left(\frac{\varphi }{M_p}\right)^n\right]^2}+1\\
  \frac{3  n^2 \xi ^2 \left(\frac{\varphi}{M_p} \right) ^2}{8 \left[1+\xi  \left(\frac{\varphi}{M_p}\right) ^2\right]^2}+1
  \end{cases} ~~;~~ \tV(\varphi) = \begin{cases}
   	\frac{V(\varphi)}{1 + \xi \left(\frac{\varphi}{M_p}\right)^n}\\
   	\frac{V(\varphi)}{\left[1+ \xi (\frac{\varphi}{M_p})^2 \right]^{\frac{n}{2}}}
   \end{cases}
   \eea
   
   	We use the following field redefinition 
   	\be
   	\frac{d\phi}{d\varphi} = F(\varphi)
   	\label{redef}
   	\ee
   	to transform the non-minimal into the action of a minimally coupled scalar field with canonical kinetic term,
   	\be
   	S_E = \int d^4 x \sqrt{- \tg} \left[\frac{M_p^2}{2}\tR  -\tg^{\mu \nu } \partial_{\mu}\phi \partial_{\nu}\phi - \tV(\phi)     \right]
   	\ee
	At this point we can integrate eq.(\ref{redef}), to find the new field in terms of the old field, and construct 
	the modified potential as a function of new field. It is clear from the above set of transformations that
	for entire range of parameter $\xi$, it is very difficult to reproduce our model. 
	However, in the regime of weak coupling $\xi << 1$, $F \sim 1$, hence we can approximately write, using eq(\ref{redef}); $\varphi \sim \phi_0 \phi$ ($\phi_0$ is some integration constant). Considering Jordan frame potential as power-law: $V(\varphi) \approx \varphi^n$, one
	gets plateau potential as
	\bea
	\tV(\phi) =
	\begin{cases}
		\frac{\lambda ~m^{4-n} \phi^n}{1 +  \left(\frac{\phi}{\phi_*}\right)^n}\\
		\frac{\lambda ~m^{4-n} \phi^n}{\left[1+  (\frac{\phi}{\phi_*})^2 \right]^{\frac{n}{2}}} ,
	\end{cases}
	\eea
where, we identify $\phi_*$ as $M_p/\xi^{\frac{1}{n}}$ for Type-I potential and $M_p/\xi^{\frac{1}{2}}$	for type-II potential. 
Therefore, in the weak coupling regime, $\xi \ll 1$ or $\phi_* > 1$, the non-minimal scalar tensor theory can give rise
to a large class of minimal cosmologies such as ours which do not belong the $\alpha$-attractor model.

\section{Validity of $1/N$ expansion for $\phi_* < M_p$}
In this section we consider $n=2$ case, as we can analytically compute the expression for the power
spectrum. Let us start by defining $x \equiv \phi/\phi_* $, and using eq.(\ref{efold}), we define the number of efolding by the following integral expression  from field value $x_k$ for which scales exit the horizon to the end of inflation with field value $x_e$, 
\bea
N_k = \int\limits_{x_{k}}^{x_{e}} \left(\frac{\phi_*}{M_p}\right)^2 \frac{V(x)}{V'(x)}~  d x .
\label{efoldx}
\eea
The $x_k$ corresponds to a scale $k$ which exits the horizon, and $x_e$ is the reduced field value at the end of inflation(~$\epsilon(x_e) =1$~). Therefore, the field value at the horizon crossing turned out to be,
\be
x_k = \left[ -1 + \sqrt{1 + f_k} \right]^{\frac{1}{2}},
\label{newparam1}
\ee
Where $f_k$ is given by
\be
f_k = 8 \frac{M_p^2}{\phi^2_*} N_k + x_e^2(2 + x_e^2).
\ee
The expression for  $x_e$ is given by
\be
x_e = \frac{2^{1/6} \left(\frac{\phi_*}{M_p}\right)^{1/3}}{3\left[ -1 + \sqrt{1 + \frac{2}{27} \left( \frac{\phi_*}{M_p}\right)^2}\right]^{1/3}} - \frac{\left[ -1 + \sqrt{1 + \frac{2}{27} \left( \frac{\phi_*}{M_p}\right)^2}\right]^{1/3}}{2^{1/6} \left(\frac{\phi_*}{M_p} \right)^{1/3}} .
\ee
In the limit $\phi_* < M_p$, which is necessary for $1/N_k$ expansion we see from the last expression that $x_e$ simply reduces to
\be
\nno
x_e \sim 2^{1/6} \left( \frac{M_p}{\phi_*} \right)^{1/3}  ~~;~~x_k \sim f_k^{1/4} .
\ee
Now the slow-roll parameters in terms of $x_k$ reduces to
\bea
\nno
\epsilon \sim \frac{2M_p^2}{\phi_*^2} \frac{1}{f_k^{3/2}}
\nno
         &=& \frac{1}{\left[8 N_k \left(\frac{M_p}{\phi_*}\right)^2 N_k + 2^{2/3} \left(\frac{M_p}{\phi_*}\right)^{4/3}\right]^{3/2}}\\
         \nno
         &\sim& \frac{1}{2^{7/2}} \frac{\phi_*}{M_p} \frac{1}{N_k^{3/2}} .
\eea
The final term in the above expression is the leading order in $N_k$ for $\phi_* < M_p$.
Similarly, the second slow-roll parameter of our interest reduces to
\bea
|\eta| \sim \frac{3}{4} \frac{1}{N_k}
\eea
All the above leading order expressions for the slow roll parameters match exactly with our general expression for $n_s$ and $r$ given in terms of $\epsilon$ and $\eta$ in Eqs.(\ref{ns-r-ns}-\ref{ns-r-R}) and it is clear that for $\phi_* \ll M_p$ $1/N_k-$expansion and slow-roll parameters are consistent. 

Another interesting limit arises for $\phi_* \gg M_p$, in this case as $\phi_*$ increases, $x_e \to 0$. Expanding eq.(\ref{newparam1}) for small $f_k( < 1)$, (which occurs when $\phi_* > 2 M_p\sqrt{2N_k}$) one arrives,
\be
x_k \sim \left(\frac{f_k}{2}\right)^{1/2},
\ee
where $f_k = 8N_k \frac{M^2_p}{\phi^2_*}$. Therefore, the slow roll parameters turn into
the following simple expressions to the leading order
\bea
\epsilon \sim |\eta| \sim 2\frac{M_p^2}{\phi_*^2} \frac{1}{x_k^2}
                     \sim \frac{1}{2 N_k}
\eea
This can be identified with the chaotic inflation limit. Indeed as we have seen before for large value of $\phi_*$, the potentials can be represented as $\phi^n$ during inflation. The results in this case are consistent with large filed models producing large value of scalar-to-tensor ratio which however, is in tension with Planck data.
Finally we also check the validity of our expansion numerically for other values of $n$. 


  \hspace{0.5cm}

\end{document}